\documentclass[]{aa}

\usepackage{graphicx}

\usepackage{txfonts}

\usepackage{subfigure}

\begin{document}

   \title{Early spectral evolution of Nova Sgr 2004 (V5114 Sgr)
     \thanks{Based on observations collected at the European Southern 
   Observatory, La Silla, Chile, Cerro Tololo Inter--American Observatory
   (CTIO), McDonald Observatory and Lick Observatory.}  }

   \subtitle{}

   \author{	A. Ederoclite\inst{1,2,3}
	  \and E. Mason\inst{2}
	  \and M. Della Valle\inst{4,5}
	  \and R. Gilmozzi\inst{2}
	  \and R. E. Williams \inst{6}
	  \and L. Germany \inst{2}
	  \and I. Saviane \inst{2}
	  \and F. Matteucci \inst{3}
	  \and B. E. Schaefer\inst{7}
	  \and F. Walter\inst{8}
	  \and R. J. Rudy\inst{9}
	  \and D. Lynch\inst{9}
	  \and S. Mazuk\inst{9}
	  \and C. C. Venturini\inst{9}
	  \and R. C. Puetter\inst{10}
	  \and R. B. Perry\inst{11}
	  \and W. Liller\inst{12}
	  \and A. Rotter\inst{13}
          }

   \offprints{A.Ederoclite}

   \institute{
      	Vrije Universiteit Brussel, 2 Pleinlaan, Brussels, Belgium
	\email{aederocl@vub.ac.be}   
   	\and
   		E.S.O. -- European Southern Observatory, Alonso de Cordova 3107,
              Casilla 91001, Santiago, Chile
         \and
             Department of Astronomy, University of Trieste, Via Tiepolo 11, 
	     Trieste, Italy
	\and
	     INAF -- Osservatorio Astrofisico di Arcetri, L.go E.Fermi 5,
	     Firenze, Italia
	\and
		Kavli Institute for Theoretical Physics, UC Santa Barbara,
		California, 93106, USA
	\and
		Space Telescope Science Institute 3700 San Martin Drive, 
		Baltimore, MD, USA
	\and
		Department of Physics and Astronomy, Louisiana State 
		University,Baton Rouge, Louisiana, 70803 USA
	\and
		Department of Physics and Astronomy, SUNY, Stony Brook, NY,
		11794-3800, USA
	\and
		Aerospace Corporation, P.O. Box 92957, Los Angeles, CA 90009-2957, U.S.A.
	\and
		University of California San Diego, 9500 Gilman Dr., La Jolla,
		CA, USA
	\and
		NASA Langley Research Center, 100 NASA Road, Hampton, VA, USA
	\and
		Isaac Newton Institute, Casilla 8-9, Correo 9, Santiago, Chile
	\and 
		Department of Astronomy and Astrophysics, Penn State University,
		525 Davey Lab, University Park, PA, USA 
             }

   \date{}

\abstract
{} 
{We present optical and near-infrared spectral evolution of
the Galactic nova V5114 Sgr (2004)
during few months after the outburst.}
{
We use multi-band photometry and line intensities derived from spectroscopy to
put constrains on the distance and the physical conditions of the ejecta of 
V5114 Sgr.
}{The nova showed a fast
decline ($t_2 \simeq 11$~days) and spectral features of \ion{Fe}{ii}
spectroscopic class. It reached $M_V = -8.7\pm 0.2$~mag at maximum
light, from which we derive a distance of $7700 \pm 700$~kpc and a
distance from the galactic plane of about 800 pc. Hydrogen and Oxygen
mass of the ejecta are measured from emission lines, leading to $\sim
10^{-6}$ and $10^{-7}$M$_\odot$, respectively.  We
compute the filling factor of the ejecta to be in the range $0.1$ -- 
$10^{-3}$. We found the value of the filling factor
to decrease with time. The same is also observed in other novae, then
giving support to the idea that nova shells are not homogeneously
filled in, rather being the material clumped in relatively higher
density blobs less affected by the general expanding motion of the
ejecta.}
{}


\maketitle

\section{Introduction}

Nova Sgr 2004 (V5114 Sgr) was independently discovered  by Nishimura (2004)
and Liller (2004) on Mar15.8~UT and Mar17.3~UT, respectively.
West (2004) provided precise
coordinates R.A. = 18\hbox{$^{\rm h}$}19\hbox{$^{\rm m}$}32\fs29, 
Decl. = $-$28\degr36\arcmin35\farcs7
(gal. coord. $l=3$\fdg$9$ $b=-6$\fdg$3$).
Early spectroscopy on Mar 18.3 UT by Della Valle et al. (2004)
confirmed  this object to be a classical nova caught near maximum light.
Here, we present photometric and spectroscopic observations taken at
McDonald Observatory, Cerro Tololo Inter-American Observatory (CTIO),
ESO-La Silla, and Lick Observatory. The spectra from ESO-LaSilla have
been obtained as part of the Target of Opportunity campaign for the
observation of Classical Novae in the Galaxy and in the Magellanic
Clouds.
The paper is organized as follows: in Sect. 2 we present the
analysis of the photometric and spectroscopic data, in Sects. 3 and
4 we analyze the evolution of the light curve and spectra of 
V5114 Sgr, in Sect. 5 we discuss the effects of interstellar
absorption. In Sect. 6 we derive the physical parameters of the nova
ejecta.  Summary and conclusions are given in Sect. 7.

   \begin{figure*}
   \centering
   \includegraphics[width=18cm, angle=-90, scale=0.8]{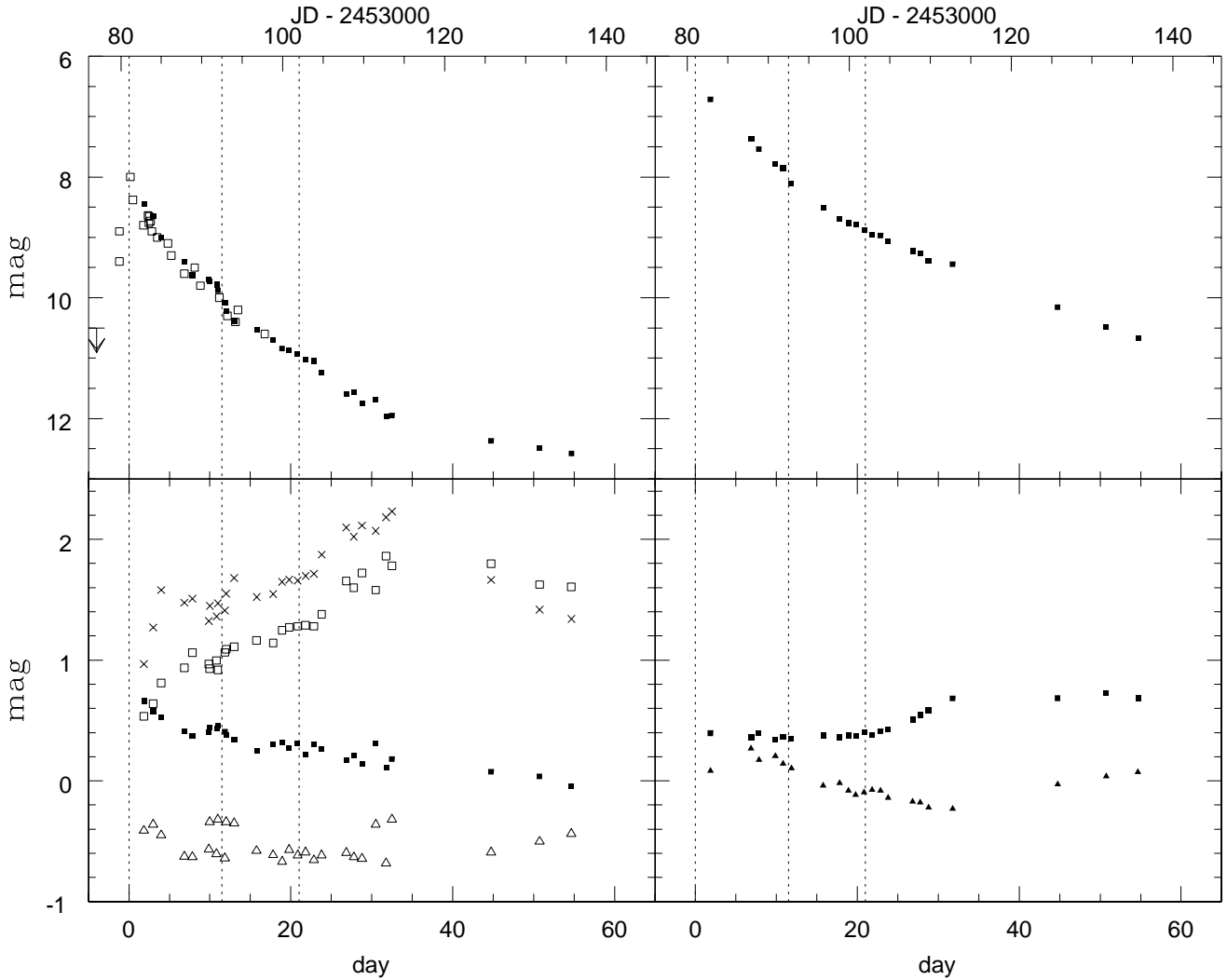}
   \caption{
\textbf{Upper left panel:}~ $V$ band light curve. Filled symbols represent
        our data points, empty symbols represent IAUC data.
	Vertical dotted lines represent maximum light, $t_2$, and $t_3$.
	Data points are affected by errors $\lesssim 0.04$~mag in all bands but
	in $U$, where the error is 0.1~mag.
        \textbf{Lower left panel: } Evolution of different colors: 
	filled squares represent $B-V$,
	empty triangles  $U-B$,
	crosses $V-I$ and empty squares $V-R$.
        \textbf{Upper right panel: } $J$ band light curve.
        \textbf{Lower right panel: } Evolution of near-infrared colors: 
	triangles represent $J-H$ and squares $H-K$.
   }
              \label{Lightcurves}
    \end{figure*}
%

\section{Observations and data reduction}
                                                                                
Optical photometry has been carried out during 8 nights in 
March and April 2004 with the 0.8-m telescope at McDonald Observatory.
Another observing run has been carried out during 14 nights in June and 
July with the Small and Moderate Aperture Research Telescope System (SMARTS)
 1.0-m telescope at CTIO.  
Optical and infrared photometry has been also carried out with the
ANDICAM dual-channel imager on the SMARTS 1.3-m at CTIO during 25 
nights from March to August.
One night of observations has been carried out with a 0.3-m telescope
equipped with an SBIG ST7E CCD camera, in Exmouth, Australia.
A log of photometric observations is given in Table~\ref{LogPhotObs}.

Spectra at maximum and during the early decline have been obtained with FEROS 
(Kaufer et al. \cite{kaufer})
with a resolution $R \sim$ 48000 and spectral range 4000 -- 9000 \AA .
Spectrophotometric standard  stars have not  been observed each night, and 
in this case the spectra have been corrected  with 
an ``average response curve''. This procedure 
can introduce an uncertainty on the flux measurement up to 50\%.
Flux-determination is affected by undetermined uncertainty because FEROS is
a fiber-fed spectrograph that was not equipped with an atmospheric distortion
corrector at the time of these observations. Indetermination is due to the 
fact that the
observations are carried out guiding on the $V$-band image of the star that is
differently displaced (due to atmospheric refraction) in the other bands.
During our analysis, fluxes were corrected in order to match the observed
magnitudes.

An independent spectroscopic follow up has been carried out with the RC 
spectrograph on the SMARTS 1.5-m telescope at CTIO.
A spectrophotometric standard star (either  LTT 4364 or Feige 110) has
been observed each night to remove the instrumental signature.  
Standard
reduction has been carried out with an author's written IDL routine.

An IR  spectrum was taken June  22, 2004 UT at  Lick Observatory using
the   Aerospace  Corporation's   Near-Infrared  and   Visible  Imaging
Spectrograph (NIRIS).  The standard star used was HR 6836.
                    
All spectra have been analyzed with the \verb+onedspec+	package in 
\verb+IRAF+. Line fluxes have been measured by the integration 
of the line profile and not by gaussian fitting. Full width at half
maximum (FWHM) of lines have been measured also via gaussian fitting but
show no significant difference from direct measure.
A complete log of our spectroscopic observation is reported in 
Table~\ref{LogObs}.

\section{Light curve}

The optical and near-infrared light curves of V5114 Sgr are shown in
Fig.~\ref{Lightcurves}.                                                                              
The light curves have been derived using both our photometric data and
photometry available in the literature (IAUC 8306, 8307, 8310). 
The $V$ light curve shows that V5114 Sgr reached $V$=8.0 mag on Mar 17.17 UT
(MJD = 53081.556).
West (2004) noted that nothing was visible at the same position in 
the red Digitized Sky Survey.
After considering that the DSS limiting magnitude is $\sim 21$ mag, we can infer that the 
outburst amplitude was $\gtrsim 13$~mag which is consistent with values observed for
other novae with about the same rate of decline (see Warner \cite{warner}).

\begin{figure}
   \centering
   \includegraphics[width=8cm]{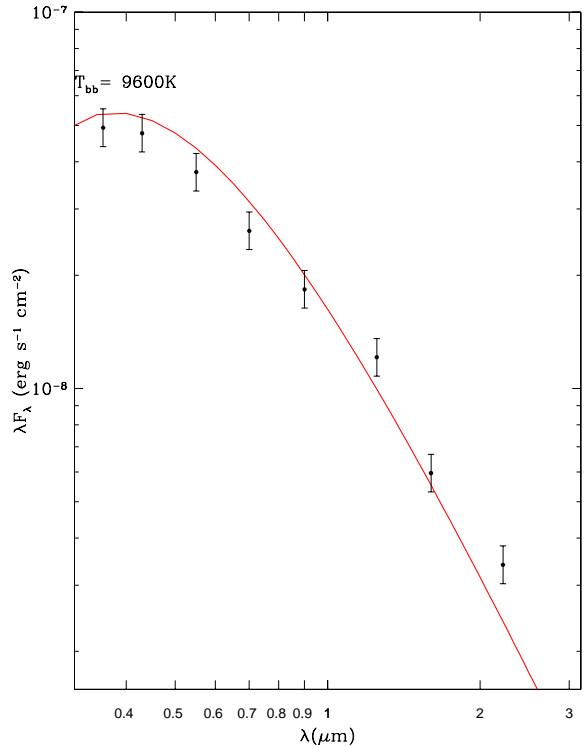}
      \caption{Spectral energy distribution of V5114 Sgr
	one day after maximum.  The magnitudes in the various bands have
	been corrected for interstellar extinction.
              }
         \label{SEDs}
   \end{figure}

The nova decreased by two magnitudes from maximum in $t_{2}=11$~days
and by three magnitudes in 
$t_{3}=21$ ~days. Adopting the maximum magnitude versus rate of
decline (MMRD) relation by Della Valle \& Livio (\cite{MDVL}), 
V5114 Sgr achieved an absolute $V$ magnitude at maximum light of $M_{V}=-8.7
\pm 0.2$ mag. Photometric properties are summarized in
Table~\ref{LCProp}.
The typical photometric errors are smaller than 0.04 mag in all bands
 but in $U \sim 0.1$ mag.
($J-H$) color changes its slope after $\sim$ day 40. This could be in
principle due to formation of dust in the ejecta but an inspection of
the $B$ and $V$ light curves rules out ``DQ Herculis'' behavior, therefore
the reddened color is likely due to to variations in intensity of
emission lines in the NIR. 
The spectral  energy distribution (SED) one day  after maximum appears
to  be well  fitted  by a  blackbody  at $T=9600$ K  as  shown in  Fig.~\ref{SEDs}.  

\begin{table}
  \caption[]{Observed photometric properties for V5114 Sgr}
  \label{LCProp}
  \begin{tabular}{p{0.4\linewidth}l r}
    \hline
    \noalign{\smallskip}
    $t_0$   &  JD =  24 53081.556 \\
    $t_2$   &  11 ~ days \\
    $t_3$   &  21 ~ days \\
    $m_{V,max}$ & 8.0 ~ mag \\
    $(B-V)_{max}$ & 0.66 ~ mag \\
    $(B-V)_{t_2}$ & 0.38 ~ mag \\
    $(B-V)_{15d}$ & 0.25 ~ mag \\
    \noalign{\smallskip}
    \hline
    \end{tabular}
  \end{table}

\section{Spectral evolution}

   \begin{figure*}
   \centering
   \includegraphics[width=18cm]{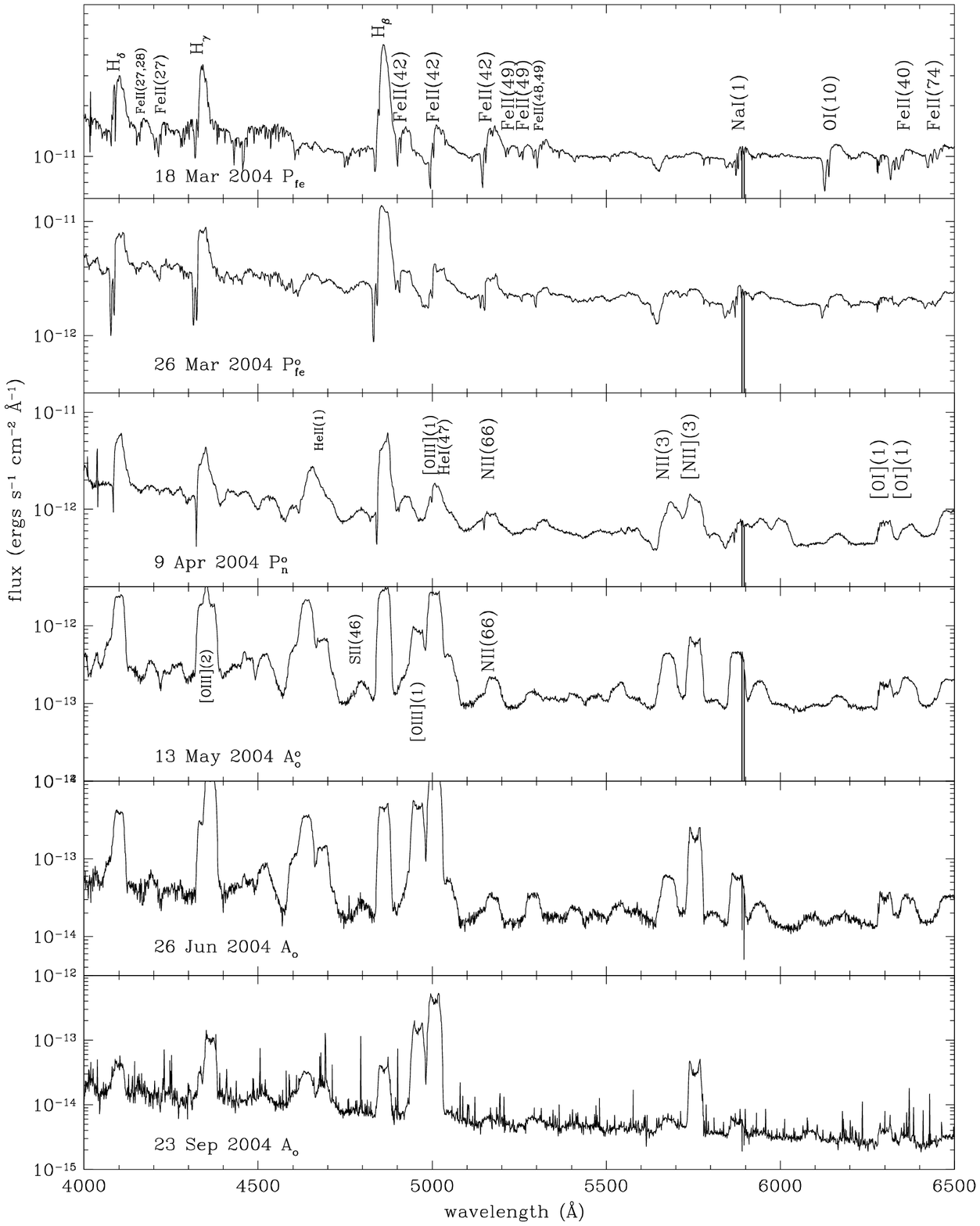} 
   \caption{Dereddened spectra of V5114 Sgr (Blue part). Fluxes are in logarithmic
                scale to show the less intense lines.}
              \label{ESOspec1}
    \end{figure*}
   \begin{figure*}
   \centering
   \includegraphics[width=18cm]{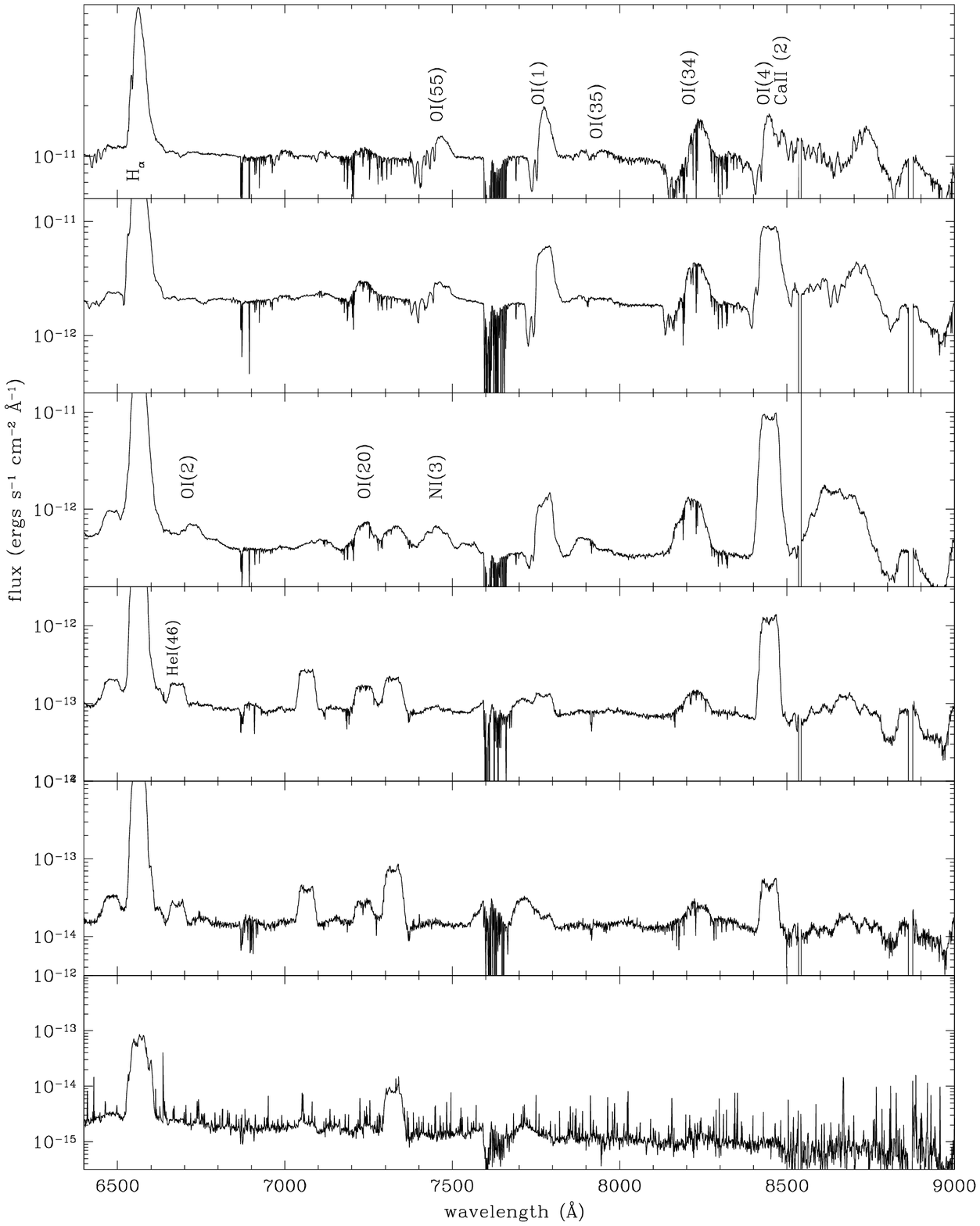}
      \caption{Dereddened spectra of V5114 Sgr (Red part). Fluxes are in logarithmic
                scale to show the less intense lines.}
              \label{ESOspec2}
	\end{figure*}

\begin{figure*}
\centering
\includegraphics[width=18cm]{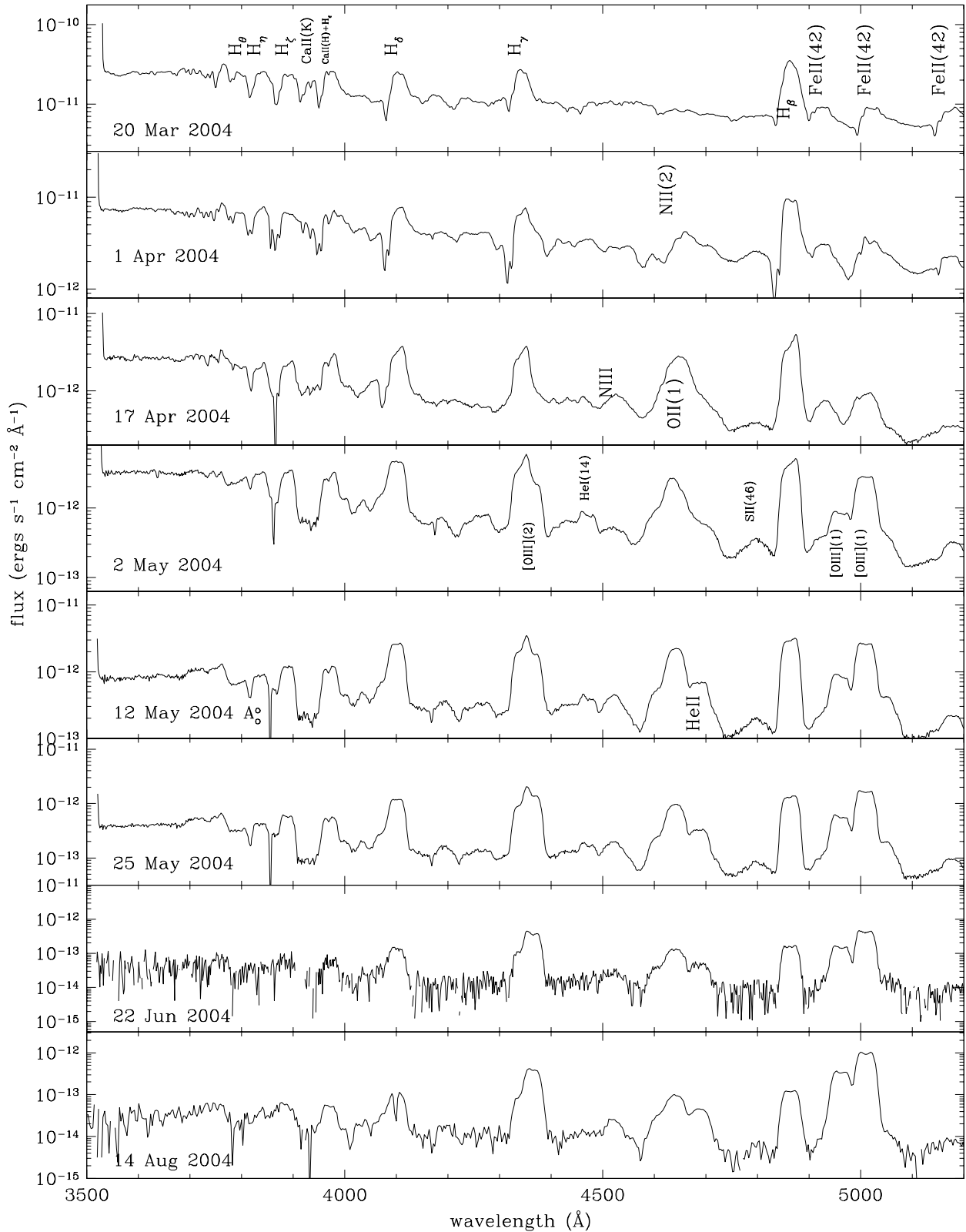}
\caption{Dereddened spectra of V5114 Sgr taken at CTIO and covering the
region 3500--5200 $\AA$.}
\label{ctiospec}
\end{figure*}

\begin{figure*}
\centering
\includegraphics[angle=-90, width=15cm]{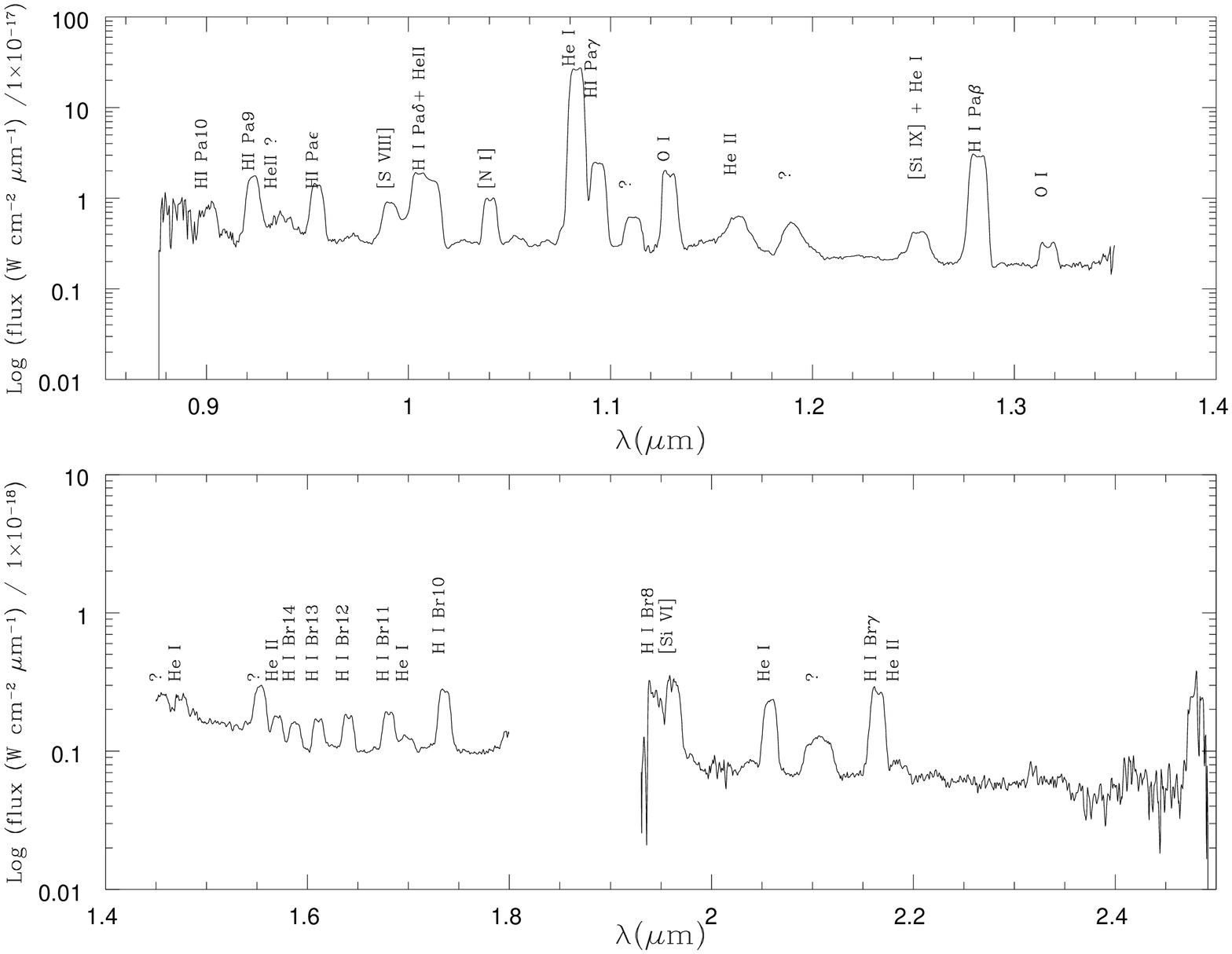}
\caption{Dereddened IR spectrum of V5114 Sgr observed with NIRIS at Lick Observatory 
 on June 22, 2004~UT. Upper panel shows the ``blue'' part of the spectrum and the lower panel 
 shows the ``red'' part.
}
	   \label{IRspec}
 \end{figure*}

Spectroscopic observations started  immediately after discovery.  Line
identification for FEROS spectra (see Fig.\ref{ESOspec1} and \ref{ESOspec2}) 
is given in Table \ref{ESOevol} 
while in Table \ref{FWlines} we show line identification for spectra taken 
at CTIO (Williams et al., {\it in preparation}).

The  first spectrum  (phase +1 \footnote{phase +1 = 1 day after maximum light}) 
was  dominated by
Balmer, \ion{Fe}{ii} and \ion{O}{i} emission lines. This behavior characterized 
the nova as a typical ``\ion{Fe}{II}''  type object, according to the 
Cerro Tololo
classification  (Williams  et  al. \cite{CTNS91}  and  \cite{CTNS94}).
P-Cyg  profiles were clearly  visible in  Balmer lines  as well  as in
\ion{Fe}{ii}, \ion{O}{i}, and  \ion{Na}{i} lines.  P-Cyg profiles were  double, 
thus suggesting the  presence of two  expanding systems  with velocities  
(obtained by averaging  of measurements  of Balmer  lines)  of $1400 \pm 
50$~km~s$^{-1}$and $850 \pm 30$~ km~ s$^{-1}$.

Eight  days after  maximum the  spectrum  was still  dominated by  low
ionization  species.  The  double  P-Cyg profiles  were still  clearly
visible and  the velocities (as  derived from both the  P-Cyg profiles
and FWHM)  were increasing.  The  emission lines started  developing a
flat topped profile.

On  April 9 (phase +23), we  observe the
4640~\AA~  emission band together  with   \ion{N}{ii} and 
\ion{N}{iii},
although  \ion{Fe}{ii} emission lines  were still present.
The \ion{O}{i} $\lambda$8446 emission line was more intense than the
H$\beta$
one and showed  a  flatter profile.

\begin{figure}
   \centering
   \includegraphics[width=8cm]{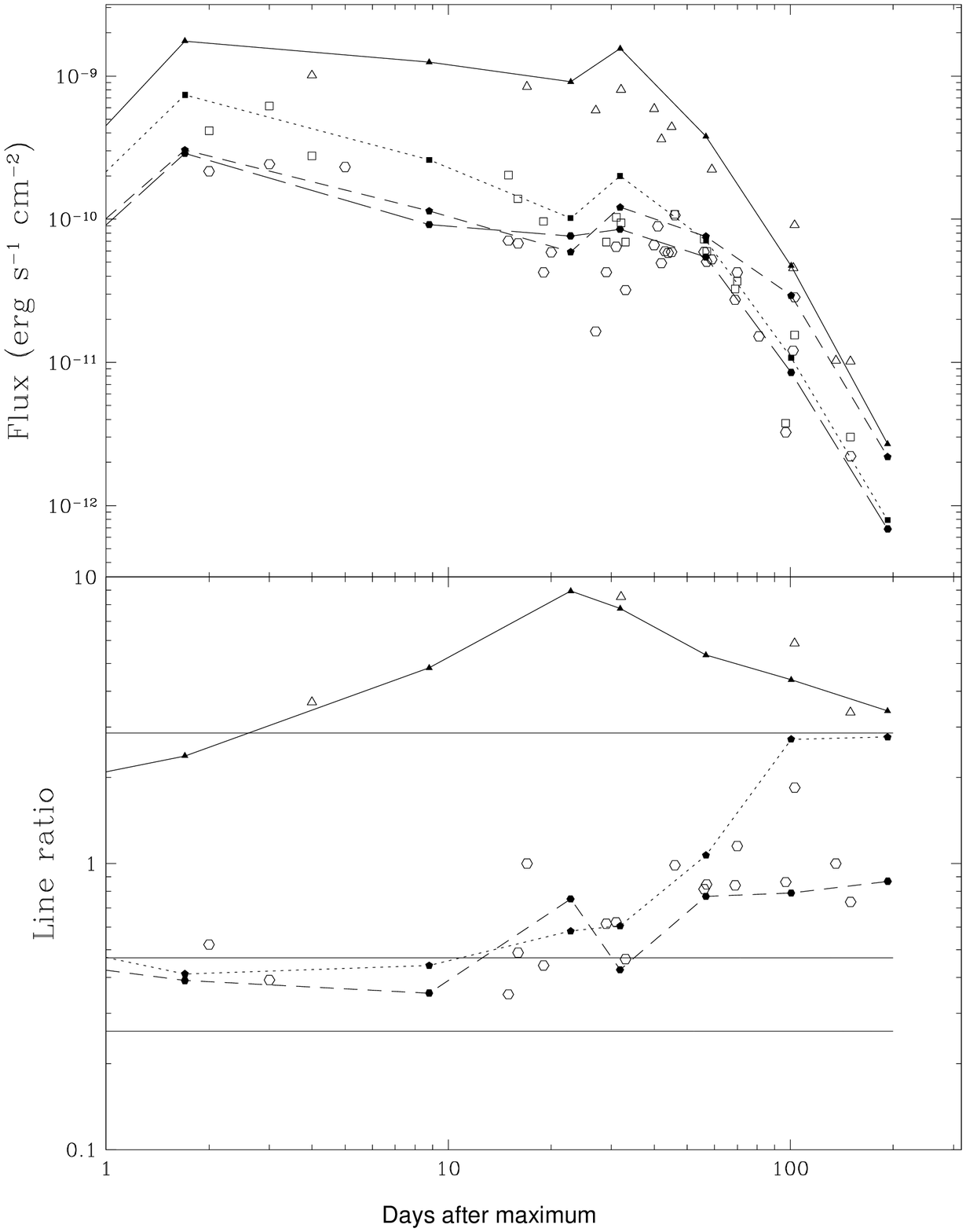}
      \caption{\textbf{Upper panel: }Evolution of Balmer lines fluxes.
Triangles represent H$\alpha$, squares
H$\beta$, pentagons
H$\gamma$
and hexagons H$\delta$. Filled symbols represent data from FEROS, empty 
symbols from the RC spectrograph at the 1.5-m telescope at CTIO. Lines
connecting the FEROS symbols are meant to show the trend of the data.
\textbf{Lower panel: }Evolution of ratios of
Balmer lines divided by the H$\beta$ flux (symbols refer to the same lines 
as in the upper panel). 
Horizontal lines represent the Balmer lines ratios for case B at 10000~K as from 
Osterbrock (1989) (from top to bottom 
H$\alpha$/H$\beta$,
H$\gamma$/H$\beta$
and H$\delta$/H$\beta$).
}
         \label{Hlines}
   \end{figure}

By April 18 (phase +32) P-Cyg profiles had disappeared and \ion{Fe}{ii}
emission lines were fading and forbidden and high excitation lines
strengthened. 
The intensity of [\ion{O}{iii}]$\lambda 4363$
indicated that V5114 Sgr entered the auroral phase, described in
Williams et al.  (\cite{CTNS91} and \cite{CTNS94}). Fluxes of
Balmer lines, that had decreased very slowly until this moment,
started to decrease faster (see Fig.\ref{Hlines}).  The FWHM of
Balmer lines reached a plateau ($2000 \pm 100$ km~s$^{-1}$).

The \ion{O}{i} $\lambda$8446 and \ion{He}{i} $\lambda$5876
lines show flat topped profiles while the
hydrogen lines have a clearly asymmetric profile (the red side being
more prominent than the blue one). At this stage the \ion{O}{i} $\lambda$8446
emission line reached its maximum intensity.

By May 13 (phase +57) the hydrogen lines turn to flat topped profiles
(like oxygen) while nitrogen lines were still rounded. It has been
noted in the past (see Payne-Gaposchkin \cite{Payne-Gaposchkin}) that
different line profiles observed at the same stage indicate that the
emission lines originate in different layers of the ejecta. 
Flat topped profiles originate from optically thin spherical shells while 
rounded profiles are related to optically thick winds.
The NIR part of the spectrum (observed only on June 22, see Fig.\ref{IRspec},
line identification given in Table \ref{OptIRfluxes}) showed prominent
Paschen and Brackett lines as well as oxygen and nitrogen lines.
Common but unknown lines (1.10 , 1.19 , 1.55 and 2.10 $\mu$m) were
present in this spectrum (see Venturini et al. \cite{Venturinietal.})
.  Tentative identifications for these lines with van Hoof's line
list\footnote{version 2.04 \textit{http://www.pa.uky.edu/ \~ \,
peter/atomic/}} are given in Table \ref{temtativeIRlines}.
Few suggested identifications have already been suggested by Rudy et al.
(\cite{Rudy02}) during analysis of lines of V723 Cas.

                                                                                
\begin{table*}
\caption{V5114 Sgr 
reddening-corrected
emission line fluxes (in ergs s$^{-1}$ cm$^{-2}$)
from FEROS spectra.}
\label{ESOevol}
\centering
\begin{tabular}{l c c c c c c c c}     
\hline\hline
Ident. wavelength & Mar18 	& Mar19 	& Mar26 	& Apr9 		& Apr18 	& May13 	& Jun26 	& Sep23\\
\hline

\ion{He}{i} 4026	& - 		& -             & 3.061-11 	& -	        & -             & 2.158E-12     & 4.457E-13	& 3.158E-13\\

[\ion{S}{ii}]  4076 	& -	  	&-	     	& -		& -    		& -     	& 1.193E-11     & 2.827E-12     & -\\
                                                                                
H$\delta$ 4101 & 3.440E-10	& 4.820E-10     & 1.925E-10 	& 2.202E-11     & 6.402E-10     & 4.634E-11     & 1.266E-11     & 2.982E-13\\

\ion{Fe}{ii}(27,28)  4173/78  & 7.030E-11 & 7.445E-11 	& 3.684E-11     & 1.603E-12      & -             & -             & -             &-\\
                                                                                
\ion{Fe}{ii}(27)  4233	& 6.490E-11     & 4.450E-11     & 2.704E-11   	& 1.258E-12    	& -             & -             & -             &-\\
                                                                                
\ion{Fe}{ii}(27)  4273	& -             & -             & 2.450E-11	& 1.247E-12     & 3.213E-12     & 2.785E-12	& -             &-\\
                                                                                
H$\gamma$ 4340 & 4.055E-10 	& 4.516E-10 	& 2.401E-10   	& 1.791E-11     & 8.476E-11     & 4.197E-11     & 1.059E-11     & 2.033E-13\\

$[$\ion{O}{iii}$]$(2) 4363 & - 		& -		& -		& -     	& -	     	& 4.012E-11     & 4.249E-11     & 6.652E-13\\
                                                                                
\ion{Fe}{ii}(27) 4417 	& -             & -             & 7.224E-12     & 3.159E-12   	& -             & -             & -             & -\\
                                                                                
\ion{He}{i}  4438   	& -     	& -             & 5.393E-12    	& -		& -             & -             & -             & - \\

\ion{He}{i}(14)  4472   & -             & -             & 2.768E-11  	& 3.384E-12 	& 5.158E-12 	& 3.990E-12     & 6.806E-13     & -\\
      
\ion{N}{iii} 4517	& - 		& - 		& - 		& 4.260E-12 	& 1.085E-11	& 5.969E-12 	& 2.921E-12	& - \\      
                                                                                
\ion{Fe}{ii}(37) 4629  & 2.678E-11    & 2.365E-11  	& -             & -		& -             & -             & -             & -\\

{\footnotesize \ion{N}{ii}(5)+\ion{N}{iii}(2)+\ion{O}{ii}(1)4601/34/49} & - & - & - 		& 2.074E-11 	& 1.243E-10 	& 6.591E-11 	& 1.843E-11 	& 2.831E-13 \\
                                                                                                                                                                                                                                                
\ion{Fe}{ii}(37) 4665	& 2.742E-11     & 1.917E-11     & 4.826E-11  	& -         	& -             & -             & -             & -\\
                                                                                
\ion{He}{ii}(1) 4686    &       - 	& - 		& - 		& - 		& - 		& 1.490E-11 	& 5.556E-12   	& 1.517E-13 \\

\ion{S}{ii}(46)  4792    & -            & -             & -             & 1.819E-12	& 2.729E-12     & 2.431E-12     & 5.134E-13     & - \\
                                                                                
H$\beta$  4861 & 8.489E-10     & 9.439E-10     & 5.244E-10     & 2.819E-11 	& 1.386E-10     & 7.492E-11     & 1.889E-11     & 2.655E-13\\
                                                                                
\ion{Fe}{ii}(42) 4924	& 1.495E-10    & 1.741E-10     & 6.124E-11     & 4.031E-12 	& 1.429E-11     & -             & -             & -\\
                                                                                
$[$\ion{O}{iii}$]$(1) 4959 & -    	& -             & -          	& -		& -             & 2.165E-11     & 1.692E-11     & 3.913E-12\\
                                                                                
$[$\ion{O}{iii}$]$(1) 5007 & -          & -             & -             & 5.972E-11	& 3.967E-11     & 7.599E-11     & 5.095E-11     & 1.191E-11\\
                                                                                
\ion{Fe}{ii}(42)  5018	& 2.633E-10     & 2.612E-10     & 1.252E-10     & -    		& -             & -             & -     	& -\\

\ion{Fe}{ii}(52)  5169  & 2.121E-10    & 2.060E-10     & 5.872E-11  	& 2.034E-11 	& 9.650E-12     & 4.177E-12     & 9.209E-13     & -\\
                                                                                                                                                                
\ion{Fe}{ii}(49)  5234  & 8.313E-11    & 3.980E-11     & 7.826E-12  	& -		& -             & -             & -             & -\\
                                                                                
\ion{Fe}{ii}(49)  5276  & 5.449E-11    & 4.539E-11     & 1.102E-11	& 3.140E-12	& -             & -             & -             & -\\

??? 5290	& -		& -		& -		& -		& -		& 1.432E-12	& 7.422E-13	& - \\
                                                                                
\ion{Fe}{ii}(49)  5317  & -            & -             & -             & 8.981E-12	& 7.669E-12     & -             & -             & -\\
                                                                                
\ion{Fe}{ii}(48)  5337  & 8.597E-11    & 1.102E-10     & 2.566E-11     & -		& -             & -             & -             & -\\
                                                                                
\ion{He}{ii}(2)   5412   & -            & -             & -             & -		& -             & 6.440E-13    	& 2.642E-13 	& -\\
                                                                                
$[$\ion{O}{i}$]$(1) 5577 & -            & -             & 1.878E-11  	& 5.771E-12	& 2.789E-12     & 6.216E-13   	& 2.222E-13     & $<$8.498E-14\\
                                                                                
\ion{N}{ii}(3) 5667/76/77/86 & -      	& -             & -             & 3.282E-11	& 1.801E-11     & 1.203E-11     & 1.983E-12 	& 1.057E-13\\

$[$\ion{N}{ii}$]$(3) 5755 & -           & 2.952E-11   & 3.544E-11   	& 4.246E-11	& 4.583E-11     & 1.703E-11     & 7.536E-12     & 9.420E-13\\
                                                                                
\ion{He}{i}(11)  5876    & -            & -             & -             & -		& 1.412E-11     & 1.090E-11     & 1.697E-12     & 8.809E-14\\

\ion{Na}{i}(1)  5890     & 2.423E-11    & 1.510E-11     & 1.978E-11     & 1.501E-11	& -             & -             &-              & -\\
                                                                                
\ion{N}{ii}(28)  5940    & -            & 1.924E-11     & 2.267E-11     & 1.761E-11	& 2.880E-12     & 3.253E-12     & 5.892E-13     & -\\
                                                                       
??? 6004 	& -		& - 		& - 		& 1.961E-11	& 8.580E-12 	& -		& -		&- \\							       
								                
\ion{O}{i}(10)  6159     & 2.423E-11    & 3.686E-11     & 1.887E-11     & 6.049E-12	& 3.257E-12     & 1.310E-12     & -             & -\\
                                                                                
\ion{Fe}{ii}(74)  6247   & 1.022E-11    & 1.440E-11     & 4.418E-12     & -		& -             & -             & -             & -\\
                                                                                
$[$\ion{O}{i}$]$ 6300   & -            & -             & 8.803E-12     & 1.152E-11	& 7.271E-12     & 2.719E-12     & 7.685E-13     & 2.507E-14 \\
                                                                                
$[$\ion{O}{i}$]$(1) 6364 & -            & -             & -             & 4.192E-12	& 3.995E-12     & 2.585E-12     & 4.408E-13     & 2.138E-14\\

\ion{Fe}{ii}(40)  6370  & 9.104E-12    & 1.834E-11     & 1.324E-11     & 4.727E-12	& -	        & -             & -             & -\\
                                                                                
\ion{N}{ii}(8)  6482     & -            & -             & 2.507E-11     & 1.983E-11	& 1.035E-11     & 4.192E-12     & 8.092E-13     & 1.470E-14\\
                                                                                
H$\alpha$  6563 & 4.634E-10    & 1.124E-9      & 1.598E-9      & 1.152E-9	& 9.801E-10      & 3.601E-10     & 5.965E-11     & 8.597E-13\\

\ion{He}{i}(46)  6678    & -            & -             & 9.063E-12   	& -		& 6.050E-12   	& 2.689E-12     & 5.309E-13     & -\\

\ion{O}{i}(2)  6726      & 9.166E-12    & 7.611E-12  	& 1.125E-11     & 1.971E-11	& 7.864E-12   	& -             & -             & -\\
                                                                                
\ion{O}{i}(21)  7002     & 1.212E-11    & 9.505E-12     & 9.552E-12     & 1.224E-12	& -	  	& -             & -             & -\\
                                                                                
\ion{He}{i}(10)  7065    & -            & -             & -             & -		& 1.023E-11     & 6.891E-12     & 1.257E-12     & 1.106E-14 \\
                                                                                
???	7113	& - 		& 9.514E-12	& -		&  -		& 3.531E-12	& -		& -		& - \\										
										
\ion{C}{ii}(3) 7231/36  & 1.195E-12  	& 4.984E-12     & 3.402E-11     & 1.951E-11	& 5.706E-12     & 3.637E-12     & 6.099E-13     & -\\
                                                                                
$[$\ion{O}{ii}$]$(2) 7319/20/30/31 & - 	& -             & -             & 1.881E-11	& 1.913E-11     & 5.491E-12     & 3.060E-12     & 9.895E-14\\
                                                                                
\ion{N}{i}(3)+\ion{O}{i}(55) 7468 + 7476 & 3.876E-11 & 7.413E-11 & 5.497E-11 	& 1.793E-11   	& 6.388E-12     & -		& -             & -\\
                          
???	7546	& - 		& -		& -		& 2.686E-12 	& 1.298E-12	& -		& -		& - \\		  
			                                                        
\ion{O}{i}(1) 7772/4/5	& 7.000E-11 	& 1.579E-10     & 2.013E-10     & 4.869E-11	& 1.936E-11     & 2.504E-12     & -             & -\\
                                                                                
\ion{Mg}{ii}(8) 7896	& - 		& - 		& - 		& 9.620E-12	& 5.092E-12	& -		& -		& - \\

\ion{O}{i}(35) 7947	& 2.086E-11     & 1.997E-11     & -             & -		& 2.286E-12	& 8.680E-13	& -             & -\\
                                                                                
\ion{O}{i}(34) 8223 	& 7.846E-11     & 1.374E-10     & 1.194E-10     & 6.634E-11	& 2.467E-11     & 3.605E-12     & 1.050E-12     & -\\
                                                                                
\ion{O}{i}(4)    8446/7 & 9.500E-11     & 1.307E-10     & 4.585E-10     & 5.154E-10	& 3.448E-10     & 4.844E-11     & 2.062E-12     & -\\

\ion{Ca}{ii} 	8498	& 2.659E-11	& 4.547E-11	& -		& -		& -		& -		& -		& - \\

\ion{Ca}{ii} 8662?	& 8.537E-12	& 1.371E-11	& -		& -		&-		& -		& -		& - \\

\hline
\end{tabular}
\end{table*}

                                                                                
\begin{table*}
\caption{V5114 Sgr 
reddening-corrected
emission line fluxes (in ergs s$^{-1}$ cm$^{-2}$)
from CTIO spectra with wavelength range larger than 2000 $\AA$
and including the H$\alpha$ region.}
\label{FWlines}
\centering
\begin{tabular}{l c c c c }    
\hline\hline
Ident. wavelength &  Mar21 	& Apr18			& Jun28		& Aug14\\
\hline
H$\epsilon$ + \ion{Ca}{ii} 3968/70& - &  -		& 5.709E-12	& 8.794E-13	\\ 
H$\delta$ 4101        & -	      & -		      & 1.823E-11     & 1.910E-12     \\
H$\gamma$ 4340 & -	      & -		      & 5.539E-11     & 1.857E-12     \\      
$[$\ion{O}{iii}$]$(2) 4363 & -		& -			& - 		& 8.853E-12 	\\
\ion{N}{iii} 4517	& -		& -			& 1.232E-12	& 2.403E-13	\\
4640		&  -		& -			& 1.277E-11	& 3.021E-12	\\
\ion{He}{ii}(1) 4686	& -		& -			& 3.334E-12	& 1.060E-12	\\
H$\beta$ 4861	& 3.295E-10	& 1.046E-10		& 1.077E-11	& 3.130E-12	\\
\ion{Fe}{ii}(42) 4924	& 4.610E-11	& 1.226E-11		& -		& -		\\
$[$\ion{O}{iii}$]$(1) 4958 & -		& -			& 1.55E-11	& 1.295E-11	\\
$[$\ion{O}{iii}$]$(1) 5007 & -		& -			& 5.69E-11	& 3.748E-11	\\
\ion{Fe}{ii}(42) 5018	& 1.751E-10	& 5.317E-11		& -		& -		\\
\ion{Fe}{ii}(52) 5169	& 9.638E-11	& 9.411E-12		& 7.794E-13	& 2.891E-13	\\
\ion{Fe}{ii}(49) 5276	& 4.055E-12	& -			& -		& -		\\
??? 5296	& -		& -			& 7.250E-13	& 3.543E-13	\\
\ion{Fe}{ii}(48) 5337	& 2.020E-11	& 5.505E-12		& -		& - 		\\
\ion{He}{ii}(2) 5412	& -		& -			& 2.154E-13	& 1.123E-13	\\
??? 	5474	& -		& -			& 9.443E-14	& -		\\
5540		& 1.432E-11	& -			& 2.917E-13	& -		\\
$[$\ion{O}{i}$]$(1)5577	& -		& -			& 1.505E-13	& $<$2.360E-13  \\
\ion{N}{ii}(3)5667/76/77/78 & 1.870E-11	& 2.218E-11		& 1.975E-12	& 4.354E-13	\\
$[$\ion{N}{ii}$]$(3) 5755 & 1.869E-11	& 4.821E-11		& 8.219E-12	& 4.477E-12	\\
\ion{He}{i}(11)	5876	& -		& 1.764E-11		& 2.054E-12	& 4.346E-13	\\
\ion{Na}{i}(1) 5890	& 2.015E-11	& -			& -		& -		\\
\ion{N}{ii}(28)	5940	& 1.520E-11	& 1.197E-12		& 6.378E-13	& 1.047E-13	\\
??? 6004	& -		& 8.634E-12		& -		& -		\\
\ion{O}{ii}(10)	6159	& 3.056E-11	& 3.412E-12		& -		& -		\\
\ion{Fe}{ii}(74) 6247	& 9.383E-12	& -			& -		& -		\\
$[$\ion{O}{i}$]$6300	& 2.858E-12	& 7.739E-12		& 6.599E-13	& 2.445E-13	\\
$[$\ion{O}{i}$]$6363 + FeII(40)6370 & 1.239E-11	& 3.287E-12	& 5.745E-13	& 1.265E-13	\\
\ion{N}{ii}(8)	6482	& 7.461E-12	& 5.879E-12		& 3.771E-13	& 1.263E-13	\\
H$\alpha$ 6563	& 1.240E-9	& 9.902E-10		& 5.747E-11	& 1.283E-11	\\
\ion{He}{i}(46) 6678	& 1.571E-12	& 3.450E-12		& 4.168E-13	& 1.020E-13	\\
\ion{O}{i}(2)	6726	& 6.578E-12	& 4.497E-12		& -		& -		\\
\ion{He}{i}(10)	7065	& -		& 1.346E-11		& 1.167E-12	& - 		\\
\ion{C}{ii}(3)	7231/36	& -		& 6.288E-12		& 5.851E-13	& - 		\\
$[$\ion{O}{ii}$]$(2) 7319/20/30/31 & -	& 2.147E-11		& 2.717E-12	& -		\\
\ion{N}{i}(3)+\ion{O}{i}(55) 7468 + 7476 & 5.187E-11 & 6.109E-12	& -		& -		\\
???	7586	& -		& -			& 1.717E-13	& -		\\
???	7630	& -		& -			& 7.579E-14	& -		\\
???	7731	& -		& -			& 9.544E-13	& -		\\
\ion{O}{i}(1)	7772/4/5 & 2.315E-10	& 2.231E-11		& -		& -		\\
\ion{Mg}{ii}(8)	7896	& -		& 5.863E-12		& -		& -	\\
\ion{O}{i}(34)	8223	& 1.302E-10	& 2.734E-11		& 3.385E-13	& -	\\
\ion{O}{i}(4) 	8446	& 2.825E-10	& 4.794E-10		& 9.237E-13	& -	\\
???	8704	& -		& -			& 6.435E-13	& -	\\
???	9030	& -		& -			& 2.261E-13	& -	\\
???	9245	& -		& -			& 9.574E-13	& -	\\
???	9556	& -		& -			& 2.905E-13	& -	\\
\hline
\end{tabular}
\end{table*}

\begin{table*}
\caption{V5114 Sgr
reddening-corrected
emission line fluxes (in ergs s$^{-1}$ cm$^{-2}$)
from CTIO spectra shown in Figure \ref{ctiospec}.}
\label{CTIOUVlines}
\centering
\begin{tabular}{l c c c c c c c c}    
\hline\hline
Ident. wavelength &  Mar20 	& Apr1		& Apr17		& May2		& May12		& May25		& Jun22		& Aug14	\\
\hline
??? 3765	& 1.553E-10	& 4.039E-11	& 1.1117E-11	& 3.114E-12	& 1.402E-11	& 1.089E-11	& 6.005E-12	& - \\
H10 3798	& 9.596E-11	& 3.257E-11	& -		& 6.295E-12	& -		& -		& -		& - \\
H9  3835	& 2.030E-10	& 8.256E-11	& 2.219E-11	& 2.242E-11	& 7.722E-12	& 6.032E-12	& -		& - \\
H8  3889	& 2.412E-10	& 9.473E-11	& 3.930E-11	& 4.155E-11	& 1.984E-11	& 1.480E-11	& 7.604E-12	& 1.146E-12\\
\ion{Ca}{ii}(K)3933	& 1.137E-10	& -		& -		& -		& -		& -		& -		& - \\
H$\epsilon$ + \ion{Ca}{ii} 3968/70& 3.719E-10 & 1.144E-10 & 4.848E-11 & 5.301E-11 & 2.259E-11	& 2.156E-11	& 5.386E-12	& 8.293E-13 \\
H$\delta$ 4101	& 3.644E-10	& 1.142E-10	& 7.861E-11	& 1.079E-10	& 8.113E-11	& 5.493E-11	& 1.760E-12	& 3.798E-12 \\
\ion{Fe}{ii}(27,28) 4173/78 & 5.184E-11	& 9.433E-12	& -		& -		& -		& -		& -		& - \\
\ion{Fe}{ii}(27) 4233	& 3.419E-11	& -		& -		& - 		& - 		& - 		& -		& - \\
H$\gamma$ 4343	& 4.241E-10	& 1.465E-10	& 8.000E-11	& 2.172E-10	& 5.656E-11	& 4.077E-11	& 1.341E-11	& 1.857E-12 \\
$[$\ion{O}{iii}$]$4363 & -		& -		& -		& 8.706E-11	& 5.334E-11	& 6.423E-11	& 4.605E-11	& 8.853E-12 \\
4422		& -		& 1.413E-11	& - 		& -		& -		& -		& -		& - \\
4468		& - 		& 2.312E-11	& -		& 1.029E-11	& -		& 2.687E-12	& -		& - \\
\ion{N}{iii}4517	& -		& -		& -		& 3.199E-12	& -		& 5.006E-12	& 2.617E-12	& 2.403E-13 \\
4640		& - 		& 1.275E-10	& 1.202E-10	& 8.310E-11	& 9.007E-11	& 5.906E-11	& 2.167E-11	& 3.021E-12 \\
\ion{He}{ii}(1)4686	& -		& - 		& -		& -		& 2.039E-11	& 1.518E-11	& 6.264E-12	& 1.060E-12 \\
H$\beta$ 4861	& 8.560E-10	& 2.517E-10	& 1.402E-10	& 1.138E-10	& 1.028E-10	& 6.833E-11	& 2.161E-11	& 3.130E-12 \\
\ion{Fe}{ii}(42) 4924	& 7.125E-11	& 3.407E-11	& 1.484E-11	& -		& -		& -		& -		& - \\
$[$\ion{O}{iii}$]$4959	& - 		& -		& -		& 1.961E-11	& 2.661E-11	& 3.037E-11	& 2.255E-11	& 1.295E-11 \\
$[$\ion{O}{iii}$]$5007	& -		& - 		& -		& 8.073E-11	& 1.008E-10	& 9.709E-11	& 6.511E-11	& 3.748E-11 \\
\ion{Fe}{ii}(42) 5018	& 1.581E-10	& 1.007E-10	& 2.767E-11	& -		& -		& -		& -		& - \\
\ion{Fe}{ii}(52) 5169	& 9.383E-11	& 2.288E-11	& 6.805E-12	& 3.897E-12	& 5.086E-12	& 2.296E-12	& -		& - \\
\hline
\end{tabular}
\end{table*}

   \begin{table}
      \caption[]{
      Reddening-corrected near infra red line fluxes 
      (in erg s$^{-1}$ cm$^{-2}$)
      observed with NIRIS at Lick Observatory on June 22, 2004, UT.}
         \label{OptIRfluxes}
         \begin{tabular}{p{0.5\linewidth}l l}
            \hline
            \noalign{\smallskip}
            wavelength &  ID     & flux \\
            \noalign{\smallskip}
            \hline
            \noalign{\smallskip}
0.9015  & \ion{H}{i}~Pa10        &       1.615E-13\\
0.9229  & \ion{H}{i}~Pa9        &       6.132E-12\\
0.9381  & ~ 		             &       1.026E-13\\
0.9545  & \ion{H}{i}~Pa$\epsilon$  &       5.604E-13\\
0.9913  & [\ion{S}{viii}]     &       4.641E-13\\
1.0049 + 1.0124  & \ion{H}{i}~Pa$\delta$+\ion{He}{ii} & 1.841E-12\\
1.0400  & [\ion{N}{i}]          &       4.165E-13 \\
1.0534  &               &       4.482E-14\\
1.0830  & \ion{He}{i}           &       1.665E-11\\
1.0938  & \ion{H}{i}~Pa$\gamma$    &       1.421E-12\\
1.1114  & ?             &       2.438E-13\\
1.1287  & \ion{O}{i}            &       1.076E-12\\
1.1626  & \ion{He}{ii} ?        &       3.708E-13\\
1.1911  & ?             &       2.826E-13\\
1.2528  & [\ion{Si}{ix}]+\ion{He}{i}    &       2.144E-13\\
1.2818  & \ion{H}{i}~Pa$\beta$     &       2.230E-12\\
1.3164  & \ion{O}{i}            &       1.044E-13\\
1.4567  &               &       5.686E-14\\
1.4760  & \ion{He}{ii}          &       5.382E-14\\
1.5528  & ?             &       1.671E-13\\
1.5719  & \ion{He}{ii}          &       3.884E-14\\
1.5881  & \ion{H}{i}~Br14        &       4.318E-14\\
1.6109  & \ion{H}{i}~Br13        &       6.792E-14\\
1.6407  & \ion{H}{i}~Br12        &       7.936E-14\\
1.6806  & \ion{H}{i}~Br11        &       1.056E-13\\
1.7002  & \ion{He}{i}           &       3.356E-14\\
1.7362  & \ion{H}{i}~Br10        &       2.157E-13\\
1.9440  &               &       2.207E-13\\
1.9621  &               &       3.045E-13\\
2.0383  &               &       1.653E-14\\
2.0581  & \ion{He}{i}           &       2.128E-13\\
2.1068  &               &       1.501E-13\\
2.1655  & \ion{H}{i}~Br$\gamma$    &       3.019E-13\\
                                                                                
            \noalign{\smallskip}
            \hline
         \end{tabular}
   \end{table}
%
   \begin{table}
      \caption[]{Tentative identification of previously unidentified NIR lines}
         \label{temtativeIRlines}
         \begin{tabular}{p{0.2\linewidth}l r r r}
            \hline
            \noalign{\smallskip}
            Observed wavelength ($\mu$ m) & Rest~frame~wavelength & 
	    Element & Transition\\
            \hline
            \noalign{\smallskip}
                                               
            1.1114   &   11112.4   &      \ion{C}{i}]   &    E1 \\
              ~      &   11112.43  &      \ion{C}{i}    &    E1 \\
              ~      &   11119.98  &      \ion{C}{i}    &    E1 \\
              ~      &   11112.    &      \ion{N}{i}    &    E1 \\
              ~      &   11114.    &      \ion{N}{i}]   &    E1 \\
              ~      &   11114.    &      \ion{N}{i}    &    E1 \\
              ~      &   11116.    &      \ion{N}{i}    &    E1 \\
              ~      &   11117.    &      \ion{N}{i}]   &    E1 \\
              ~      &   11117.    &      \ion{N}{i}]   &    E1 \\
              ~      &   11115.3   &      \ion{N}{i}I   &    E1 \\
              ~      &   11116.00  &      \ion{N}{i}I]  &    E1 \\
              ~      &   11120.10  &      \ion{N}{i}I   &    E1 \\
              ~      &   11108.6   &      \ion{N}{i}II] &    E1 \\
              ~      &   11116.48  &      \ion{O}{ii}   &    E1 \\
              ~      &   11116.27  &      \ion{O}{ii}I] &    E1 \\
                                                        
            \hline
             1.1911  &    11905.516   &    \ion{He}{ii}  &    E1  \\
             ~       &    11907.4     &    \ion{C}{i}]   &    E1  \\
             ~       &    11915.35    &    \ion{C}{i}    &    E1  \\
             ~       &    11905.6     &    \ion{C}{ii}   &    E1  \\
             ~       &    11906.1     &    \ion{C}{iv}   &    E1  \\
             ~       &    11908.3     &    \ion{C}{iv}   &    E1  \\
             ~       &    11907.      &    \ion{N}{i}]   &    E1  \\
             ~       &    11910.2     &    \ion{N}{i}]   &    E1  \\
             ~       &    11910.2     &    \ion{N}{i}    &    E1  \\
             ~       &    11911.005   &    \ion{O}{i}]   &    E1  \\
             ~       &    11911.090   &    \ion{O}{i}]   &    E1  \\
             ~       &    11914.54    &    \ion{O}{ii}I] &    E1  \\
                                                                                
             \hline
             1.4567  &   14566.14    &    \ion{C}{i}]   &    E1  \\
             ~       &   14564.9     &    \ion{C}{ii}   &    E1  \\
             ~       &   14568.7     &    \ion{C}{ii}   &    E1  \\
             ~       &   14560.7     &    \ion{C}{iii}  &    E1  \\
             ~       &   14560.7     &    \ion{C}{iii}  &    E1  \\
             ~       &   14563.964   &    \ion{O}{i}]   &    E1  \\
             ~       &   14564.136   &    \ion{O}{i}]   &    E1  \\
                                                 
             \hline
                                                                                
             1.5528 & 15514.13    &    \ion{He}{i}   &    E1\\
             ~      & 15524.5     &    \ion{C}{i}    &    E1\\
             ~      & 15527.71    &    \ion{C}{i}]   &    E1\\
             ~      & 15531.3     &    \ion{C}{i}    &    E1\\
             ~      & 15528.085   &    \ion{N}{i}]   &    E1\\
             ~      & 15524.371   &    \ion{O}{i}    &    E1\\
             ~      & 15524.388   &    \ion{O}{i}]   &    E1\\
                                                                                
             \hline
             2.0383 & 20373.28   &     \ion{He}{ii}  &    E1\\
             ~      & 20379.9    &     \ion{N}{i}    &    E1\\
             ~      & 20389.     &     \ion{N}{i}   &	 E1\\
             ~      & 20387.63   &     \ion{O}{ii}   &    E1\\
             ~      & 20389.2    &     \ion{O}{ii}   &    E1\\
             \hline
             2.1068 & 21067.     &     \ion{C}{iii}  &    E1\\
             ~      & 21067.     &     \ion{C}{iii}  &    E1\\
             ~      & 21068.     &     \ion{C}{iii}  &    E1\\
             ~      & 21068.     &     \ion{C}{iii}  &    E1\\
             ~      & 21061.     &     \ion{C}{iv}   &    E1\\
             ~      & 21067.     &     \ion{N}{i}]   &    E1\\
             ~      & 21067.2    &     \ion{N}{i}    &    E1\\
             ~      & 21067.6    &     \ion{N}{i}    &    E1\\
             ~      & 21068.0    &     \ion{N}{i}    &    E1\\

            \noalign{\smallskip}
            \hline
         \end{tabular}
   \end{table}

In September the spectrum was dominated by \ion{O}{i} $\lambda \lambda$ 
4959--5007 lines.
All lines showed saddle-shaped profiles. 
The \ion{O}{i} $\lambda$8446 line had almost disappeared. 
H$\alpha$ was clearly strongly blended with \ion{N}{ii}.

   \begin{figure}
   \centering
    \includegraphics[angle=-90, width=8cm, scale=.5]{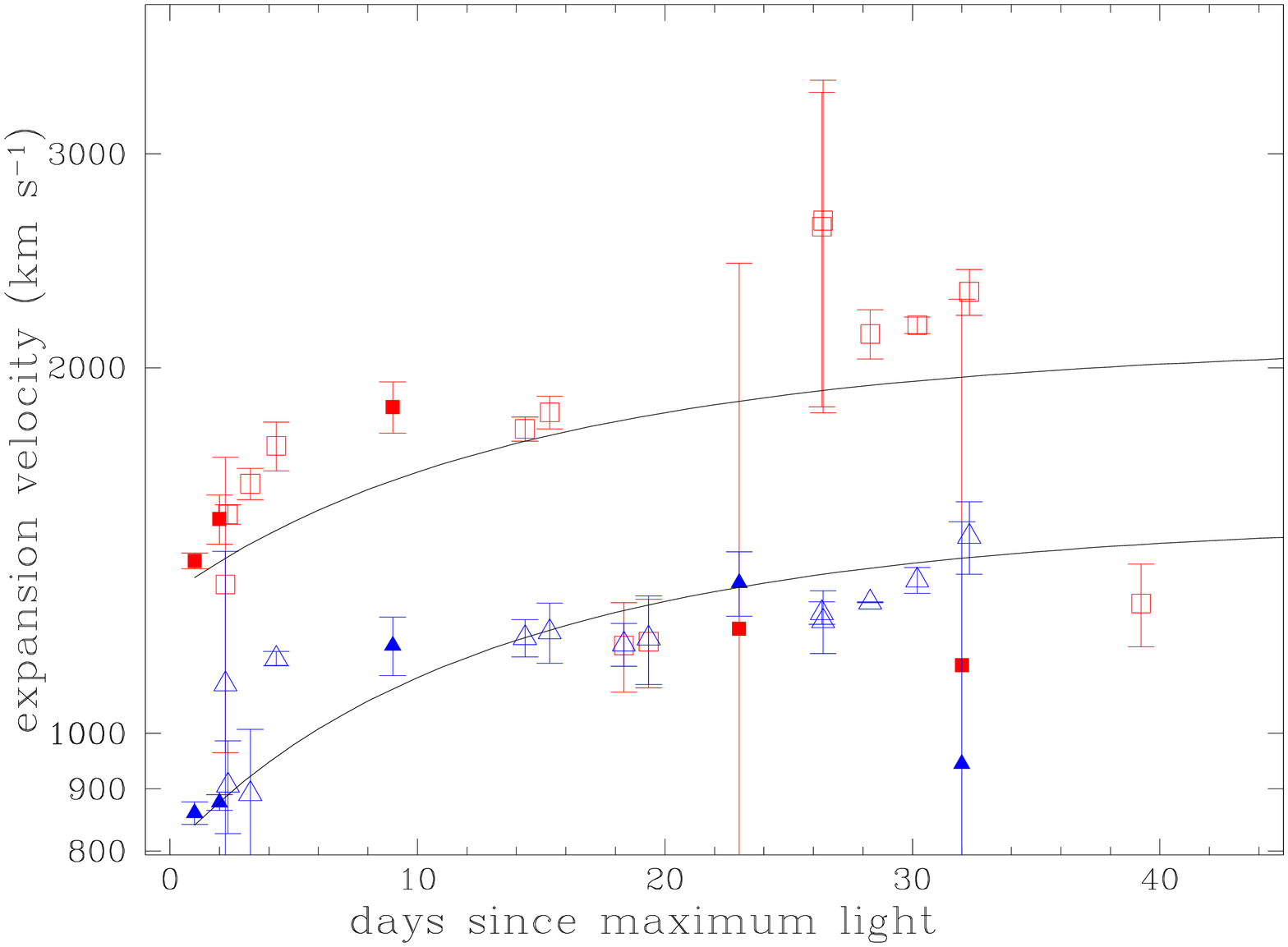}
      \caption{
	Velocities of the ejecta. Measures are derived as average of
        P-Cyg profiles of Balmer lines.
	Squares refer to the faster absorption system and triangles to the
	slower one. Filled symbols refer to FEROS observations and empty sybols
	to CTIO observations.
        Continuous lines are exponential description of the evolution
        (see text).
                }
         \label{velocities}
   \end{figure}
The overall evolution of expansion velocities measured from P-Cyg
absorption is shown in Fig.\ref{velocities}. Cassatella et al.
(\cite{Cassatella}) have shown that the evolution of P-Cyg profiles as
measured in the UV of Nova Cyg 1992 can be modelled with an exponential
law $v(t) = v_\infty - (v_0 - v_\infty) e^{-t/\tau} $, where $v(t)$ is
the velocity $t$ days after maximum, $v_0$ is the expansion velocity
at maximum light, $v_\infty$ is the asymptotic velocity and $\tau$ is
a time scale similar to $t_3$.  
The best fit to the data gives
$v_\infty = 2100 \pm 50$ ~km s$^{-1}$ and $1500 \pm 50$ ~km s$^{-1}$,
$v_0 = 1300 \pm 50$~ km s$^{-1}$ and $800 \pm 50$ km s$^{-1}$ and
$\tau = 18\pm 1$~day.  Both $v_\infty$ and velocity measurements
based on the FWHM of emission lines during the nebular stage are
important quantities that should be considered in view of future
observations aimed at determining the distance to V5114 Sgr, via
nebular expansion parallax.

\section{Reddening and distance}

 We have used several methods to estimate the reddening of 
V5114 Sgr.  This piece of information is crucial to measure the
distance to the object. The results of our various estimates of the
interstellar reddening are summarised in Table~\ref{ReddTabl}.
   \begin{table}
      \caption[]{Measured values for reddening of V5114 Sgr.}
         \label{ReddTabl}
	\begin{tabular}{l l l}
            \hline
            \noalign{\smallskip}
            method & reference &  $E_{B-V}$      \\
            \noalign{\smallskip}
            \hline
            \noalign{\smallskip}
	$(B-V)_{max}$	& van~ den~ Bergh~ \&~ Younger~(\cite{vdBY}) &  0.43 \\
	$(B-V)_{t_2}$	& van~ den~ Bergh~ \&~ Younger~ (\cite{vdBY}) &  0.40 \\
	\ion{K}{i} $\lambda$7699& Munari ~\&~ Zwitter (\cite{Munari})   &  0.45\\
	\ion{O}{i} emission line ratios	& Rudy~ et~ al.~ (\cite{Rudy91})	&  1.23 \\
	Dust maps	&	Schlegel~ et~ al.~ (\cite{Schlegel}) &  0.58 \\
	\ion{He}{ii}$\lambda\lambda$10124,4686 & Williams
	\cite{Williams94}  &  0.57 \\
	\ion{H}{I}$\lambda\lambda$10049,4861 & Williams
	(\cite{Williams94}) & 0.65 \\
	\ion{He}{i}$\lambda\lambda$10830,4471 &  Osterbrock
	\cite{Osterbrock}  & 0.9 \\
	\ion{He}{i}$\lambda\lambda$5876,4471 &  Robbins
	(\cite{robbins}) & 0 \\
            \noalign{\smallskip}
            \hline
         \end{tabular}
   \end{table}

We could not use the commonly used NaD lines because they are saturated. 
Nevertheless it has
been shown by Munari \& Zwitter (\cite{Munari}) that alternatively to 
these lines, the \ion{K}{i} $\lambda$7699 line can be used. 
The advantage to use this line is that it has a less steep curve
of growth, thus giving the opportunity to measure higher reddenings.
After applying this method to V5114 Sgr, we found $E_{B-V} = 0.45 $~mag.
Colors  at   maximum  are  a   widely  used  empirical   indicator  of
interstellar reddening.  Van Den Bergh \& Younger (\cite{vdBY}) showed
that novae at  maximum have an intrinsic $B-V$ of  $0.23 \pm 0.06$ ~
mag.  The  same authors showed  also that novae, two  magnitudes below
maximum, have  an intrinsic  $ B-V =  -0.02 \pm 0.04$~mag.   As from
Table~\ref{LCProp},  V5114 Sgr  had $(B-V)_{\textrm{max}} = 0.66$  ~mag and
$(B-V)_{t_2} = 0.25$~mag.  This leads to $E_{B-V} = 0.43 \pm 0.06$~mag
and  $ E_{B-V}  =  0.40 \pm  0.04$~mag, respectively.   
From Schlegel et al. (\cite{Schlegel}) dust maps of the Milky Way,
we have derived $ E_{B-V}  = 0.58$~mag. 
Relations based on emission-line ratios are considered the best ones
(see Williams \cite{Williams94}), unfortunately they can be used only
at very late stages when the nova is in the optically thin phase. As an
example, if we compute the reddening through \ion{O}{i} $\lambda$8446 and 
\ion{O}{i} $\lambda$13164 
(following Rudy et al. 1991) we find $E_{B-V}=1.2$~mag, which is very different
from the reddening estimates derived with other methods (see Table~\ref{ReddTabl}). 
Williams (\cite{Williams94}) suggested that if  the 
\ion{He}{ii} $\lambda\lambda$4686,10124 and the 
\ion{H}{i} $\lambda\lambda$4861,10049 are optically thin, then they can be used 
to measure the absorption to the nova via:
\begin{displaymath}
E_{(B-V)_{\ion{He}{ii}}}=1.01 \log (4.1 F_{10124}/F_{4686})
\end{displaymath}
and
\begin{displaymath}
E_{(B-V)_{\ion{H}{i}}}=1.08 \log (17 F_{10049}/F_{4861}) \quad .
\end{displaymath}
The observed flat-topped profiles suggest that these lines likely originate in
the ejected shell and therefore they may be optically thin. 
Therefore
we can derive $E_{B-V}=0.57$~mag and
$E_{B-V}=0.65$~mag, respectively. 
A last attempt has been carried out making use of the fact that the 
\ion{He}{i} triplet ratios $\lambda\lambda$5876,4471 and 
$\lambda\lambda$10830,4471 are  rather unsensitive to density.
Robbins (\cite{robbins}) showed that $\frac{F_{5876}}{F_{4471}}\sim 2.9$ for  
different values of density and temperature, while,
from Osterbrock (\cite{Osterbrock}), $\frac{F_{10830}}{F_{4471}} = 4.4$ for 
typical values of temperature and  density.
From the observed \ion{He}{i} $\lambda\lambda$5876,4471
lines we find zero extinction. This value is unlikely to be correct and
Ferland (\cite{ferland})
adviced against the use of these two lines because of the small
baseline and the faintness of the \ion{He}{i} $\lambda$4471 line. On the other
hand, the \ion{He}{i} $\lambda\lambda$4471,10830 lines point towards
$E_{B-V}=0.9$~mag thus supporting rather high extinction.
In
the following we adopt $E_{B-V}=0.6\pm0.3$~mag.

After assuming the maximum magnitude obtained in Section 3 with the MMRD 
($M_V = -8.7 \pm 0.2$~mag) and the
average absorption derived above, we  derive a distance to V5114 Sgr of
$9000\pm 900$~pc. A complementary estimate of  the distance has  been obtained
using  the Buscombe-DeVaucouleurs  relation  (all novae show  the  same
magnitude   15   days   after   maximum)   as   from   Capaccioli   et
al. (\cite{Capaccioli}). This second estimate leads to $6000\pm1200$~pc.
After taking the average weight, we get $d = 7700 \pm 700$~pc.
For $l=3$\fdg$9$
and 
$b=-6$\fdg$3$, we find 
the nova  to be located $770 \pm 70$~pc above the  galactic plane

\section{Physical parameters}

Analysis of dereddened line fluxes  is the only way to derive physical
parameters of nova ejecta (e.g. masses and temperatures).
Since ejecta are still evolving toward the nebular stage, the 
line ratios are not the ones expected from atomic transition
probabilities. For example, looking at Fig.\ref{Hlines} the 
H$\alpha$/H$\beta$ ratio converges toward the theoretical value only
after phase $\sim 100$.

A different explanation (also at later stages) is invoked for the case of
[\ion{O}{i}] $\lambda\lambda$6300, 6364. This is a well known example of lines
that does not respect the theoretical ratio $\sim$ 3:1. 
Williams (\cite{Williams94}) interpreted this as due to large optical 
depth in the $6300$\AA~ line, and showed that
the optical depth of that line can be derived from 
\begin{equation}
\label{opticaldepthform}
\frac{j_{6300}}{j_{6364}}
=\frac{1-e^{-\tau}}{1-e^{-\tau/3}}
\end{equation}
Following this argument, we find $\tau_{6300}$ to be in the range 
1.6 -- 6.1, in good agreement with the values exhibited by other novae.
The optical depth, together with the ratio of [\ion{O}{i}] 
$\lambda\lambda$6300, 5577 can be used to determine electron temperature of
the zones of the ejecta where [\ion{O}{i}] lines are formed through the formula:

\begin{equation}
  T_{e}=\frac{11200}{\log[43\tau/(1-e^{-\tau})\times 
      F_{\lambda6300}/F_{\lambda5577}]}
\end{equation}
We find $T_e$ to be between $3700$~K and $6000$~K.

The knowledge of the optical depth and electron temperature allows us to estimate
the mass of oxygen in the ejecta using the $6300$ \AA~ line:
\begin{equation}
M_{\ion{O}{i}}=152 d ^2 _{\mathrm{kpc}}
\exp { \frac{22850}{T_e} } \times 10^{1.05E(B-V)}
\frac{\tau}{1-e^{-\tau}}F_{\lambda6300}\mathrm{M}_\odot
\end{equation} 
We find $M_{\ion{O}{i}}= 1.9 \times 10^{-5} -   2.4 \times 10^{-7}$ M$_\odot$.

Electron densities can be determined adopting the temperatures computed 
above
and [\ion{O}{iii}] line ratios as in Osterbrock (\cite{Osterbrock}) 
\begin{equation}
	\frac{j_{4959} + j_{5007}}{j_{4363}} = 7.73 
	\frac{e^{3.29 \times 10^{-4}/T_e}}{1+4.5 \, 10^{-4} \frac{N_e}{T_e^{1/2}}}
\end{equation}
The values we obtain are in the range $10^7$ -- $10^9$~cm$^{-3}$, close to
the upper limit of the critical densities to give rise to nebular and auroral
lines. This is an indication that these lines likely arise from regions
characterized by a relatively high density.

Hydrogen mass can be derived following Mustel \& Boyarchuk
(\cite{MustelBoyarchuk}). The H$\alpha$ line is given by
\begin{equation}
\label{Halphaflux}
j_{H\alpha} = g_\alpha \times n^2 _e \times \epsilon \times V 
\end{equation}
where $g_\alpha$ is the emission coefficient, $n_e$ is the electron
density, $\epsilon$ is the so-called ``filling factor" (a measure of
the clumpiness of the ejecta) and $V$ is the volume that can be
expressed as 
\begin{equation}
\label{volume}
V = 4 \pi \times R^2 \times \delta  
\end{equation}
where the radius $R$
can be obtained by simple expansion $v_{\mathrm{exp}} \times \Delta t$ and
$\delta = R ~ v_{\mathrm{ther}} / v_{\mathrm{exp}} $ is assumed to be representative of
the thickness of the shell.  
Deriving $v_{\mathrm{exp}}$ from the
FWHM of Balmer lines ($\sim 2000$~km 
s$^{-1}$) and 
$v_{\mathrm{ther}} = \sqrt{3kT/m_p}$ (
for $T_e \sim 5 \, 10^3$~K, 
$v_{\mathrm{ther}} \sim 10$~km s$^{-1}$) it follows
that $\delta \sim 0.01\times R$
(cfr.  $v_{\mathrm{ther}}/v_{\mathrm{exp}} \sim 0.05$ for GK Per,  Mustel \& Boyarchuck 1970).
On the other hand this is true only for the case of a very thin shell
instantaneously ejected. In fact, the shell ejection process occurs on a much
longer time scale, of the order of $t_3$. Therefore the thickness of the
shell can be computed, as order of magnitude, by the ratio between $t_3 \times
v_{\mathrm{exp}}$ and $t_{\mathrm{neb}} \times v_{\mathrm{exp}}$~,
being $t_{\mathrm{neb}}$ the time needed by the nova to reach the nebular stage,
 i.e. $\delta \sim 20\mathrm{days}/100\mathrm{days}$ = $0.2 \times R$. 
This is consistent with spectroscopic measurements obtained by Della Valle et al.
(1997) for FH Ser ($\delta=0.5 \times R$) and Humason (1940) for DQ Her 
($\delta = 0.5 \times R$). 

Solving Eq.\ref{Halphaflux} for $\epsilon$, 
\begin{equation}
 \epsilon = \frac{j_{H_\alpha}  d^2}{ g_\alpha  n_e^2 V }
\end{equation}
we find that the filling factor spans a  range of $7.1 \times
10^{-2}$ -- $7.9 \times 10^{-4}$. Finally, we can determine
from 
\begin{equation}
M_H = n_e  m_H 4 \pi R^3 \epsilon \,\, 0.2
\end{equation}
 values of
$M_H$ to be $\sim 3.0 \times 10^{-5} - 1.1 \times 10^{-6} $M$_\odot$.

The derived physical parameters are summarized in Tables \ref{Physpars1} and
\ref{Physpars2}.
\begin{table*}
\caption{Physical parameters for V5114 Sgr: \ion{O}{i} $\lambda\lambda6300/6363$ 
line ratios, optical depth in $\lambda6300$ ($\tau _{6300}$), electron 
temperature, \ion{O}{i} mass, [\ion{O}{iii}] line ratios and electron densities.}
\label{Physpars1}
\centering
\begin{tabular}{c c c c c c c c c c c}
\hline\hline
Date   & $j_{6300}/j_{6363}$	&$\tau$& $T_e$(K)& $M_{\ion{O}{i}}$(M$_\odot$) & 
$\frac{j_{4959}+j_{5007}}{j_{4363}}$ & $N_e$(cm$^{-3}$)\\
\hline
Jun26	& 1.74	& 2.10	& 4390	& 3.02E-6 & 1.60 & 1.28E9 \\
Jun28 	& 1.15	& 6.09	& 3659	& 1.87E-5 & 2.37 & 3.53E9 \\
Aug14  & 1.93	& 1.60  & 5743 & 2.36E-7 & 5.69 & 7.01E8 \\
Sep26  & 1.17   & 5.69  & 6022 & 5.73E-7 & 23.79 & 1.30E7 \\
\hline
\end{tabular}
\end{table*}

\section{Summary and conclusions}

V5114 Sgr is an \ion{Fe}{ii} nova that occurred in the bulge of the Milky
Way.  The rate of decline characterizes V5114 Sgr as a borderline
object between the fast ($t_2 < 12$days) and slow ($t_2 > 12$ days)
classes (Della Valle \& Livio \cite{MDVL98}).  These authors have
shown that He/N and \ion{Fe}{ii}b (i.e. {\it \ion{Fe}{ii} broad}) novae belong 
to the fast class and are preferentially concentrated towards the galactic
disc, i.e. at small $z$ above the galactic plane ($z<200$~pc), while
\ion{Fe}{ii} novae belong to both slow and fast classes and  are observed
both in the disk and in the bulge, extending up to $\sim 1$~kpc. 
V5114 Sgr located at about 0.8~kpc above the galactic plane in the
direction of the galactic bulge is an \ion{Fe}{ii} nova which does not
represent an exception to this scenario (Della Valle et al. \cite{MDVetal92}).

Spectroscopic observations showed a dramatic change in the overall
appearance of the spectrum of the nova at about 30 days after maximum
light (in coincidence with the entrance in the ``auroral phase''):
permitted lines start fading at a different rate, P-Cyg profiles
disappear, and velocities reach a ``plateau'' phase and $U-V$ and
$J-K$ colors increase. The lack of detection of [\ion{Ne}{iii}] line hints
for a ``standard'' evolution of the nebular spectrum (see Williams et
al. 1994).

The very high values of optical depth in the \ion{O}{i}$\lambda6300$ suggest
very high densities for the zones of the ejecta where these lines are
formed.  We have derived the filling factor in the range $\sim 7.1 
\times 10^{-2}$ 
to $7.9 \times 10^{-4}$.  Comparing these values with other values
reported in the literature (see Table \ref{filfactab}), two facts emerge:
a) the filling factors in nova ejecta are definitely smaller than 1,
likely close to 0.1, during the early stages; b) these values decrease
by 1--3 orders of magnitude with time.  
This fact indicates that the volume of the expanding shell (computed with
Eq.\ref{volume}) increases with time more rapidly than the volume actually 
occupied by most of the ejected material. In other words, the decreasing 
trend exhibited by the filling factor suggests that the ejected matter tends 
to remain clumped in sub-structures having higher density than the average 
density characterizing the expanding shell. This is consistent with the
$[$\ion{O}{i}$]$ 
$\lambda\lambda$6300, 6364 ratio
$< 2$ (Table \ref{Physpars1}), which suggests that $[$\ion{O}{i}$]$ lines should 
be formed in very dense small blobs of neutral material embedded within the 
ionized shell (see paragraph 6). 
However, recently Williams \& Mason (2006) proposed an alternative
interpretation of this behavior, assuming that the $[$\ion{O}{i}$]$ lines arise in
regions of high magnetic field, and their intensity and profile are
modified by Quadratic Zeeman Effect. 
\begin{table}
\caption{Physical parameters for V5114 Sgr: 
 $\epsilon$ and the hydrogen mass.}
\label{Physpars2}
\centering
\begin{tabular}{c c c}
\hline\hline
Date	& $\epsilon$	& $M_H$ (M$_\odot$)\\
\hline
Jun26  & 7.08E-2 & 3.02E-5	 \\
Jun28  & 7.04E-2 & 2.91E-5	 \\
Aug14  & 3.97E-3 & 6.49E-6	 \\
Sep26  & 7.92E-4 & 1.10E-6	 \\
\hline
\end{tabular}
\end{table}

Computed oxygen and hydrogen masses are in the ranges $ 
1.9 \times 10^{-5} - 2.4 \times 10^{-7} $M$_\odot$ and $ 3.0 \times 10^{-5} - 
1.1 \times 10^{-6} $M$_\odot$. This high mass ratio is close to the upper limit
for classical novae shown in Warner (\cite{warner}).

\begin{table}[h]
	\caption[]{Comparison between filling factors in literature.}
	\label{filfactab}
\begin{tabular}{l l l}
	\hline
	Nova	&	Filling factor value	&	Reference	\\
	\hline
	T Pyx		&	$10^{-2}$--$10^{-5}$ & Shara et al. (\cite{Shara}) \\
	Nova Vel 1999	&	0.1		& 	Della Valle et
	al.(\cite{MDVPDW}) \\
	Nova SMC 2001	&	0.1 -- $10^{-3}$&	Mason et al. 
	(\cite{Mason2005}) \\
	Nova LMC 2002	&	$10^{-2}$ -- $10^{-4}$&	Mason et al. 
	(\cite{Mason2005}) \\
	\hline
	{\bf V5114 Sgr } & {\bf $0.1$ -- $10^{-3}$}	& 
	{\bf \textit{this paper} }\\ 
	\hline
\end{tabular}
\end{table}

\begin{acknowledgements}
The authors are indebted to Pierluigi Selvelli and Chris Sterken for their critical reading of the manuscript.
They also thank the anonymous referee and Steve Shore, who helped
improving the presentation.

This work has been partly supported by "IAP P5/36" Interuniversity Attraction
Poles Programme of the Belgian Federal Office for Scientific, Technical and 
Cultural Affairs.

\end{acknowledgements}

\appendix

\section{Logs of the observations}
~

\begin{table*}
\caption{Measured magnitudes during photometric campaigns (described in text).}
  \label{LogPhotObs}
\centering
\small
\begin{tabular}{l c c c c c c c c}
\hline
$JD$	& $U$	& $B$ 	& $V$ 	& $R$	& $I$ 	& $J$	& $H$	& $K$	\\
\hline\hline

2453083.25 &  8.70 &  9.11 &  8.45 &  7.91 &  7.48 & 6.72 &  6.63 &  6.24 \\ 
2453084.00 &  8.87 &  9.23 &  8.65 &  8.01 &  7.38 & --   & --    & -- \\ 
2453084.25 &  8.74 &  9.24 &  8.73 &  8.07 &  7.55 & --   & --    & -- \\ 
2453084.50 &  8.87 &  9.23 &  8.65 &  8.01 &  7.38 & -- & -- & -- \\ 
2453085.00 &  9.08 &  9.53 &  9.00 &  8.19 &  7.42 & -- & -- & -- \\ 
2453085.50 &  9.08 &  9.53 &  9.00 &  8.19 &  7.42 & -- & -- & -- \\ 
2453088.25 &  9.19 &  9.81 &  9.40 &  8.46 &  7.93 &  7.37 &  7.10 &  6.74 \\ 
2453089.25 &  9.38 & 10.01 &  9.63 &  8.57 &  8.12 &  7.54 &  7.37 &  6.97 \\ 
2453091.25 &  9.54 & 10.10 &  9.70 &  8.74 &  8.38 & 7.79 &  7.59 &  7.25 \\ 
2453091.50 &  9.83 & 10.17 &  9.73 &  8.80 &  8.28 & -- & -- & -- \\
2453092.25 &  9.61 & 10.22 &  9.78 &  8.79 &  8.42 &  7.86 &  7.72 &  7.35 \\ 
2453092.50 & 10.02 & 10.34 &  9.88 &  8.96 &  8.41 & -- & -- & -- \\ 
2453093.25 &  9.85 & 10.49 & 10.08 &  9.02 &  8.67 &  8.11 &  8.01 &  7.66 \\ 
2453093.50 & 10.26 & 10.60 & 10.22 &  9.13 &  8.67 & -- & -- & -- \\ 
2453094.50 & 10.38 & 10.73 & 10.39 &  9.28 &  8.71 & -- & -- & -- \\ 
2453097.25 & 10.20 & 10.78 & 10.53 &  9.37 &  9.01 &  8.52 &  8.56 &  8.18 \\ 
2453099.25 & 10.39 & 11.00 & 10.70 &  9.55 &  9.15 &  8.70 &  8.72 &  8.36 \\ 
2453100.50 & 10.49 & 11.16 & 10.84 &  9.59 &  9.19 &  8.76 &  8.84 &  8.46 \\ 
2453101.25 & 10.57 & 11.14 & 10.87 &  9.60 &  9.20 &  8.78 &  8.89 &  8.52 \\ 
2453102.25 & 10.63 & 11.24 & 10.93 &  9.65 &  9.27 &  8.88 &  8.97 &  8.57 \\ 
2453103.25 & 10.64 & 11.23 & 11.02 &  9.73 &  9.32 &  8.95 &  9.02 &  8.64 \\ 
2453104.25 & 10.69 & 11.34 & 11.05 &  9.77 &  9.33 &  8.97 &  9.05 &  8.64 \\ 
2453105.25 & 10.90 & 11.51 & 11.24 &  9.86 &  9.37 &  9.06 &  9.20 &  8.77 \\ 
2453108.25 & 11.17 & 11.77 & 11.60 &  9.94 &  9.50 &  9.23 &  9.40 &  8.89 \\ 
2453109.25 & 11.14 & 11.77 & 11.56 &  9.96 &  9.54 &  9.26 &  9.44 &  8.89 \\ 
2453110.25 & 11.25 & 11.90 & 11.76 & 10.04 &  9.64 &  9.39 &  9.61 &  9.03 \\ 
2453112.00 & 11.64 & 12.00 & 11.69 & 10.11 &  9.62 & -- & -- & -- \\ 
2453113.25 & 11.39 & 12.07 & 11.95 & 10.10 &  9.77 &  9.44 &  9.68 &  8.99 \\ 
2453114.00 & 11.81 & 12.13 & 11.95 & 10.17 &  9.72 & -- & -- & -- \\ 
2453126.25 & 11.85 & 12.44 & 12.37 & 10.57 & 10.70 & 10.16 & 10.19 &  9.50 \\ 
2453132.25 & 12.03 & 12.53 & 12.50 & 10.87 & 11.08 & 10.48 & 10.45 &  9.72 \\ 
2453136.25 & 12.10 & 12.53 & 12.58 & 10.97 & 11.24 & 10.67 & 10.59 &  9.91 \\ 
2453156.27 & \multicolumn{8}{c}{{\it unfiltered} 12.36} \\
2453186.25 & 14.35 & 14.11 & 13.67 & 13.20 & 13.60 & -- & -- & -- \\ 
2453187.00 & 14.39 & 14.16 & 13.68 & 13.25 & 13.66 & -- & -- & -- \\ 
2453188.25 & 14.37 & 14.17 & 13.68 & 13.26 & 13.58 & -- & -- & -- \\ 
2453189.25 & 14.44 & 14.19 & 13.77 & 13.40 & -- & -- & -- & -- \\ 
2453190.25 & 14.46 & 14.21 & 13.74 & 13.34 & 13.76 & -- & -- & -- \\ 
2453191.25 & 14.51 & 14.25 & 13.76 & 13.40 & 13.81 & -- & -- & -- \\ 
2453192.00 & 14.53 & 14.26 & 13.77 & 13.42 & 13.82 & -- & -- & -- \\ 
2453193.00 & 14.56 & 14.31 & 13.82 & 13.46 & 13.82 & -- & -- & -- \\ 
2453194.25 & 14.59 & 14.33 & 13.83 & 13.50 & 13.90 & -- & -- & -- \\ 
2453195.25 & 14.62 & 14.33 & 13.83 & 13.53 & 13.95 & -- & -- & -- \\ 
2453196.25 & 14.66 & 14.40 & 13.86 & 13.58 & 13.99 & -- & -- & -- \\ 
2453197.25 & 14.69 & 14.39 & 13.87 & 13.61 & 14.00 & -- & -- & -- \\ 
2453198.25 & 14.68 & 14.43 & 13.88 & 13.62 & 14.04 & -- & -- & -- \\ 
2453200.00 & 14.78 & 14.50 & 13.92 & 13.70 & 14.08 & -- & -- & -- \\ 

\hline

\end{tabular}

\end{table*}

\begin{table*}
  \caption[]{Log of the spectroscopic observations for V5114 Sgr.}
  \label{LogObs}
  $$
  \begin{tabular}{l l l l l}
    \hline
    \noalign{\smallskip}
    Date   &  Instrument & Exp.time     & Wavelength  & Resolution \\
    (UT)    &       ~       &       (sec)   &       range (\AA)  & (or scale)\\
    \noalign{\smallskip}
    \hline
    \noalign{\smallskip}
    Mar18.3 &       FEROS   &       240     &       4000--9000    & 48000   \\ 
    Mar19.3 &       FEROS   &       400     &       4000--9000    & 48000   \\ 
    Mar19.3 &       SMARTS  &       360     &       4000--5000    & 0.77    \\ 
    Mar19.3 &       SMARTS  &       360     &       4000--5000    & 0.77    \\ 
    Mar20.3 &       SMARTS  &       270     &       3500--5300    & 1.5     \\ 
    Mar21.4 &       SMARTS  &       120     &       4800--9500    & 5.6     \\ 
    Mar22.4 &       SMARTS  &       600     &       3900--4500    & 0.6     \\ 
    Mar26.4 &       FEROS   &       632     &       4000--9000    & 48000   \\
    Apr1.4  &       SMARTS  &       360     &       3500--5300    & 1.5     \\
    Apr2.3  &       SMARTS  &       360     &       4000--4900    & 0.77    \\
    Apr3.3  &       SMARTS  &       300     &       5900--7700    & 1.5     \\
    Apr5.3  &       SMARTS  &       360     &       3500--5300    & 1.5     \\
    Apr6.4  &       SMARTS  &       360     &       3900--4550    & 0.56    \\
    Apr9.4  &       FEROS   &       900     &       4000--9000    & 48000   \\
    Apr13.3 &       SMARTS  &       360     &       5600--7000    & 1.1     \\
    Apr13.4 &       SMARTS  &       360     &       3800--4550    & 0.56    \\
    Apr13.4 &       SMARTS  &       360     &       4050--4750    & 0.56    \\
    Apr15.4 &       SMARTS  &       540     &       4000--5000    & 0.77    \\
    Apr16.3 &       SMARTS  &       720     &       3500--5300    & 1.48    \\
    Apr17.3 &       SMARTS  &       360     &       3500--5300    & 1.48    \\
    Apr18.4 &       FEROS   &       900     &       4000--9000    & 48000   \\
    Apr18.4 &       SMARTS  &       300     &       4800--9500    & 5.6     \\
    Apr18.4 &       SMARTS  &       180     &       4800--9500    & 5.6     \\
    Apr19.4 &       SMARTS  &       480     &       3800--5600    & 1.48    \\
    Apr26.2 &       SMARTS  &       450     &       5600--6950    & 1.1     \\
    Apr26.4 &       SMARTS  &       540     &       3850--4550    & 0.56    \\
    Apr27.3 &       SMARTS  &       450     &       3870--4550    & 0.56    \\
    Apr28.2 &       SMARTS  &       600     &       3870--4550    & 0.56    \\
    Apr28.4 &       SMARTS  &       600     &       5650--7000    & 1.1     \\
    Apr29.2 &       SMARTS  &       600     &       3870--4550    & 0.56    \\
    Apr30.2 &       SMARTS  &       450     &       3870--4550    & 0.56    \\
    May1.2  &       SMARTS  &       450     &       3870--4550    & 0.56    \\
    May1.4  &       SMARTS  &       450     &       5600--7000    & 1.1     \\
    May2.2  &       SMARTS  &       360     &       3500--5300    & 1.48    \\
    May12.4 &       SMARTS  &       360     &       3500--5300    & 1.48    \\
    May13.3 &       FEROS   &       1500    &       4000--9000    & 48000   \\
    May13.4 &       SMARTS  &       360     &       3500--5300    & 1.48    \\
    May14.3 &       SMARTS  &       360     &       3500--5300    & 1.48    \\
    May15.1 &       SMARTS  &       450     &       5650--7000    & 1.1     \\
    May15.4 &       SMARTS  &       450     &       3900--4550    & 0.56    \\
    May25.3 &       SMARTS  &       720     &       3500--5300    & 1.48    \\
    May26.3 &       SMARTS  &       720     &       4000--5000    & 0.77    \\
    Jun6.2  &       SMARTS  &       720     &       4050--4750    & 0.56    \\
    Jun22   &       NIRIS   &       ?       &       4500--25000   & ?       \\
    Jun22.2 &       SMARTS  &       360     &       3500--5300    & 1.48    \\
    Jun26.2 &       FEROS   &       3600    &       4000--9000    & 48000   \\
    Jun27.1 &       SMARTS  &       450     &       5650--7000    & 1.1     \\
    Jun27.4 &       SMARTS  &       720     &       3850--4550    & 0.56    \\
    Jun28.1 &       SMARTS  &       360     &       4800--9600    & 5.67    \\
    Jul15.1 &       SMARTS  &       600     &       3550--5300    & 1.48    \\
    Jul31.2 &       SMARTS  &       540     &       5650--7000    & 1.10    \\
    Aug14.2 &       SMARTS  &       600     &       3450--6900    & 2.88    \\
    Sep26.2 &       FEROS   &       7200    &       4000--9000    & 48000   \\
    \noalign{\smallskip} 
    \hline
    \end{tabular}
  $$
  \end{table*}

\end{document}